\documentclass[runningheads]{llncs}

\usepackage[T1]{fontenc}
\usepackage{graphicx}
\usepackage{amsmath}
\usepackage{xcolor}
\usepackage{amssymb}
\usepackage{algorithm}
\usepackage[noend]{algpseudocode}
\usepackage{booktabs}
\usepackage{multirow}

\DeclareMathOperator*{\argmax}{\arg\!\max}

\begin{document}

\title{Don't Hash Me Like That: Exposing and Mitigating Hash-Induced Unfairness in Local Differential Privacy}

\titlerunning{Exposing and Mitigating Hash-Induced Unfairness in LDP}

\author{
    Berkay Kemal Balioglu\thanks{These authors contributed equally to this work.} \and 
    Alireza Khodaie\textsuperscript{*} \and 
    M. Emre Gursoy
}
\authorrunning{Balioglu, Khodaie, and Gursoy}

\institute{Department of Computer Engineering, Koç University, Istanbul, Türkiye\\
\email{\{bbalioglu23, akhodaie22, emregursoy\}@ku.edu.tr}
}

\maketitle   
\begin{abstract}
Local differential privacy (LDP) has become a widely accepted framework for privacy-preserving data collection. In LDP, many protocols rely on hash functions to implement user-side encoding and perturbation. However, the security and privacy implications of hash function selection have not been previously investigated. In this paper, we expose that the hash functions may act as a source of unfairness in LDP protocols. We show that although users operate under the same protocol and privacy budget, differences in hash functions can lead to significant disparities in vulnerability to inference and poisoning attacks. To mitigate hash-induced unfairness, we propose Fair-OLH (F-OLH), a variant of OLH that enforces an entropy-based fairness constraint on hash function selection. Experiments show that F-OLH is effective in mitigating hash-induced unfairness under acceptable time overheads. 


\keywords{Privacy \and local differential privacy \and protocol fairness \and inference attacks \and poisoning attacks \and privacy technologies and mechanisms}
\end{abstract}

\section{Introduction}

Local Differential Privacy (LDP) has become a widely adopted framework for collecting user data while providing strong privacy guarantees \cite{cormode2018privacy,xiong2020comprehensive,yang2023local}. By perturbing each user's data locally before transmission, LDP protocols eliminate the need for a trusted data collector. Many LDP mechanisms were proposed in the literature, and several prominent deployments of LDP exist in the real world, e.g., in Chrome, Apple iOS, and Microsoft Windows \cite{ding2017collecting,erlingsson2014rappor}.

While LDP protocols offer formal privacy guarantees, recent research has highlighted that several privacy and security risks remain. Adversaries can utilize protocol outputs to make predictions regarding users' true values via inference attacks \cite{arcolezi2023risks,gadotti2022pool,gursoy2022adversarial}, or inject manipulated reports to skew aggregate statistics estimated by the data collector via poisoning attacks \cite{mga,cheu2021manipulation,li2023fine,wu2022poisoning}. To implement encoding and perturbation, many LDP protocols such as OLH rely on hash functions. However, the security and privacy implications of hash function selection, e.g., its impacts on inference and poisoning attacks, have not been investigated.

In this paper, for the first time (to the best of our knowledge), we expose that hash functions can act as a source of unfairness in LDP. We show that even when users operate under the same protocol and privacy budget $\varepsilon$, differences in hash functions can lead to significant disparities in vulnerability to inference and poisoning attacks. In particular, users whose hash functions exhibit low uniformity (i.e., yielding skewed distributions of hash outputs) are more negatively affected by inference attacks and cause more severe damage via poisoning attacks. We experimentally demonstrate these traits using three subpopulations (High-ENT, Low-PIS, and High-PIS), which represent balanced, unfairly advantaged, and unfairly disadvantaged groups according to their hash functions. Results with five datasets, a state-of-the-art inference attack (BIA \cite{gursoy2022adversarial}), and poisoning attack (MGA \cite{mga}) show that: (i) Low-PIS users are 1.5-2 times more vulnerable to BIAs, and (ii) High-PIS users' impact via MGAs can be 3-4 times higher than others.

To mitigate hash-induced unfairness, we propose a fair variant of the well-known OLH protocol, called Fair-OLH (F-OLH). The main idea of F-OLH is to enforce each user's hash function to have sufficiently high fairness, measured in terms of entropy. In F-OLH, the user keeps drawing hash functions from the underlying hash family until a suitable hash function that satisfies the fairness threshold $\rho$ is found. The remainder of user-side perturbation and server-side estimation in F-OLH is similar to OLH. 

We show that F-OLH is effective in mitigating hash-induced unfairness. As $\rho$ gets stricter, different subpopulations' vulnerabilities to BIAs converge to one another. Furthermore, the utility loss caused by MGAs is substantially reduced in F-OLH compared to OLH. On the other hand, since F-OLH enforces the user to draw hash functions until a suitable hash function is found, it brings an increased time cost. We experimentally establish that this time cost is related to the domain size, privacy budget $\varepsilon$, as well as the threshold $\rho$, since finding hash functions which satisfy stricter $\rho$ is more time-consuming. Experiment results show that although the execution time of F-OLH is larger than OLH, it still remains acceptable, and replacing OLH by F-OLH in a practical deployment is unlikely to cause a noticeable problem.

\textbf{Contributions.} In summary, our main contributions include:
\vspace{-3pt}
\begin{itemize}
\item We expose the problem of hash-induced unfairness in LDP for the first time, using a state-of-the-art inference attack (BIA) and poisoning attack (MGA). 
\item To mitigate hash-induced unfairness, we propose a fair variant of OLH, a popular and well-known LDP protocol, called Fair-OLH (F-OLH).
\item We provide extensive results demonstrating that F-OLH reduces disparate impacts of BIAs and MGAs, thereby achieving better fairness, without introducing prohibitive time costs. 
\end{itemize}

\section{Background and Preliminaries} \label{sec:Background}

\subsection{Local Differential Privacy} \label{sec:LDP}

Local Differential Privacy (LDP) is a widely used notion for safeguarding users' privacy in data collection. Unlike central differential privacy (DP) which assumes the existence of a trusted data collector, LDP ensures that each user locally perturbs their data on their own device before sharing the perturbed version with the data collector. In a typical LDP setting, a population of users, denoted by $\mathcal{U}$, interacts with a server that aggregates the received data. Each user $u \in \mathcal{U}$ holds a true value $v_u \in \mathcal{D}$, where $\mathcal{D}$ represents the domain of possible values. To satisfy LDP, each user applies a randomized mechanism $\Psi$ to perturb their $v_u$ before sending the result to the server. The server can estimate population-level aggregate statistics from the received responses, but cannot reconstruct any individual's true value.

\begin{definition}[$\varepsilon$-LDP]
A randomized mechanism $\Psi$ satisfies $\varepsilon$-LDP if and only if, for any two possible inputs $v_1, v_2$ $\in$ $\mathcal{D}$, the following condition holds:
\begin{equation}
\forall y \in Range(\Psi): ~~~~ \frac{\text{Pr}[\Psi(v_1) = y]}{\text{Pr}[\Psi(v_2) = y]} \leq e^{\varepsilon}
\end{equation}
where $Range(\Psi)$ represents the set of all possible outputs of $\Psi$. 
\end{definition}

LDP is typically implemented using well-known LDP protocols such as GRR, BLH, OLH, RAPPOR, OUE, and SS \cite{arcolezi2022improving,cormode2021frequency,gursoy2022adversarial,wang2017locally}. In many of the protocols, $v_u$ is encoded into a lower-dimensional artifact such as a bitvector, integer, or a single bit. For example, RAPPOR and OUE encode $v_u$ into a bitvector \cite{erlingsson2014rappor,wang2017locally}; BLH and OLH encode $v_u$ into a bit and integer, respectively \cite{gursoy2022adversarial,wang2017locally}. To facilitate this encoding, hash functions are commonly used. This paper focuses on the privacy and security impacts associated with the choice of these hash functions. In particular, we will utilize the Optimized Local Hashing (OLH) protocol to demonstrate the impacts, since OLH is a popular and state-of-the-art protocol that is commonly used in recent works \cite{wang2020locally,yang2023local}. 

\textbf{Optimized Local Hashing (OLH)} was introduced by Wang et al.~\cite{wang2017locally}. It utilizes a set of hash functions $\mathcal{H}$, where each hash function $H \in \mathcal{H}$ maps values from $\mathcal{D}$ to an integer in the range $[0, g-1]$. The default value of $g$ is derived as $g = e^{\varepsilon} + 1$ in \cite{wang2017locally}, which we also use in the rest of this paper.

In OLH, each user $u$ randomly selects a hash function $H_u$ from $\mathcal{H}$ and computes the integer $x_u = H_u(v_u)$. Then, the perturbation mechanism $\Psi_{\text{OLH}}$ perturbs $x_u$ into $x_u' \in [0,g-1]$ using the following probabilities:
\begin{equation} \label{eq:OLH_perturb}
\Pr[x_u' = i] =
\begin{cases}
\frac{e^{\varepsilon}}{e^{\varepsilon} + g - 1}, & \text{if } x_u = i, \\
\frac{1}{e^{\varepsilon} + g - 1}, & \text{otherwise}.
\end{cases}
\end{equation}
The user sends the tuple $\langle H_u, x_u' \rangle$ to the server. After collecting tuples $\langle H_u, x_u' \rangle$ from users $u \in \mathcal{U}$, the server begins the aggregation and estimation phase. To perform estimation for some $v \in \mathcal{D}$, the server first finds $Sup(v)$, which is equal to the number of user tuples satisfying $x'_u = H_u(v)$. The server then finds the estimated frequency of $v$, denoted by $\tilde{f}(v)$, using the formula:
\begin{equation}
\tilde{f}(v) = \frac{(e^{\varepsilon} + g - 1) \cdot (g \cdot Sup(v) - |\mathcal{U}|)}{(e^{\varepsilon} - 1) \cdot (g - 1) \cdot \mathcal{|U|}}
\end{equation}

\subsection{Bayesian Inference Attack} \label{sec:BIA}

There are several recent security and privacy attacks on LDP protocols, such as reidentification attacks \cite{murakami2020toward}, pool inference attacks \cite{gadotti2022pool}, and poisoning attacks \cite{cheu2021manipulation,li2023fine,wu2022poisoning}. Among them, we utilize two applicable attacks to demonstrate hash-induced unfairness in LDP: one privacy attack (Bayesian Inference Attack) and one poisoning attack (Maximal Gain Attack). 

Bayesian Inference Attack (BIA) was first proposed in \cite{gursoy2022adversarial}, and later extended and used in \cite{arcolezi2025revisiting,arcolezi2023risks,gursoy2024longitudinal}. Having observed one or more protocol outputs originating from the user's true value $v_u$, the attacker aims to infer the most likely value that resulted in the observed outputs via Bayesian inference. More formally, let $\mathcal{O}_u$ denote the protocol outputs that the adversary observed from $u$. In the case of OLH, if data is collected once, then $\mathcal{O}_u = \langle H_u, x_u' \rangle$. If data is collected multiple times longitudinally, then we write $\mathcal{O}_u = \{ \langle H^1_{u},x^1_{u} \rangle, \langle H^2_{u},x^2_{u} \rangle, ..., \langle H^n_{u},x^n_{u} \rangle \}$. Armed with $\mathcal{O}_u$, the goal of the adversary is to predict $v_u$. Denoting the adversary's prediction by $v_u^p$, according to the Bayes theorem:
\begin{align} \label{eq:conf}
v^p_u &= \argmax_{\hat{v} \in \mathcal{D}} ~\text{Pr}[\hat{v}|\mathcal{O}_u] = \argmax_{\hat{v} \in \mathcal{D}} ~\frac{\text{Pr}[\mathcal{O}_u|\hat{v}] \cdot \text{Pr}[\hat{v}]}{\text{Pr}[\mathcal{O}_u]} \\ &= \argmax_{\hat{v} \in \mathcal{D}} ~\text{Pr}[\mathcal{O}_u|\hat{v}] \cdot \text{Pr}[\hat{v}] \\
&\propto \argmax_{\hat{v} \in \mathcal{D}} ~\text{Pr}[\mathcal{O}_u|\hat{v}]
\end{align}
From the definition of the OLH protocol, we know that for all pairs of values $v_j,v_k \in \mathcal{D}$ such that $H_u(v_j) = H_u(v_k)$, it holds that: $\text{Pr}[x^{i}_u|v_j] = \text{Pr}[x^{i}_u|v_k]$. In other words, the two values $v_j,v_k$ that are both within the preimage set of the hash function $H_u$ are equally likely. Second, consider that a certain $x^{i}_u$ was reported to the server, and we divide $\mathcal{D}$ into two disjoint subsets $\mathcal{D}_1, \mathcal{D}_2$ such that $\mathcal{D}_1 = \{ v | v \in \mathcal{D}, H^{i}_u(v) = x^{i}_u \}$ and $\mathcal{D}_2 = \mathcal{D} \setminus \mathcal{D}_1$. Then, due to Equation \ref{eq:OLH_perturb}, it holds that: $\text{Pr}[v_u \in \mathcal{D}_1] > \text{Pr}[v_u \in \mathcal{D}_2]$. 

\begin{algorithm}[!t]
\caption{Bayesian Inference Attack on OLH}
\label{alg:BIA}
\begin{algorithmic}[1]
\Require $\mathcal{D}$, $\mathcal{O}_u = \{ \langle H^1_{u},x^1_{u} \rangle, ..., \langle H^n_{u},x^n_{u} \rangle \}$
\Ensure Predicted value $v_u^p$
\For{$v \in \mathcal{D}$}
\State Initialize $score(v) = 0$
\EndFor
\For{$i \in [1,n]$}
    \For{$v \in \mathcal{D}$}
        \If{$H_u^{i}(v) = x_u^{i}$}
            \State $score(v) = score(v) + 1$ 
        \EndIf
    \EndFor
  \EndFor
\State Find $v_u^p = \argmax\limits_{v} ~score(v)$
\State \Return $v_u^p$
\end{algorithmic}
\end{algorithm}

Combining these observations, the BIA for OLH can be implemented as shown in Algorithm \ref{alg:BIA}. The algorithm initializes the score of each $v \in \mathcal{D}$ as 0. Then, for each observed $\langle H^i_{u},x^i_{u} \rangle$, the algorithm computes the preimage of $x^i_{u}$ according to $H^i_{u}$, i.e., the subset of values in the domain which would result in $x^i_{u}$ when hashed using $H^i_{u}$. The scores of each of those values are incremented by 1. Finally, the attacker selects the highest scoring value and returns it as the predicted $v_u^p$. This strategy conforms to the aforementioned observations because (i) those values that are in the preimage set are more likely than those that are not, and (ii) all values in the preimage set are equally likely; therefore, their scores are incremented by +1. 

\vspace{-6pt}
\subsection{Maximal Gain Attack} \label{sec:MGA}

LDP protocols are vulnerable to poisoning attacks in which attackers inject manipulated data to distort the estimated statistics on the server side (e.g., $\tilde{f}(v)$). Poisoning attacks can be untargeted, aiming to degrade the overall accuracy of frequency estimation, or targeted, focusing on inflating or reducing the estimated frequencies of attacker-chosen items. 

Maximal Gain Attack (MGA) is a seminal poisoning attack introduced in \cite{mga}. In MGA, there exist a set of target items $T = \{\bar{v}_1, \bar{v}_2, ..., \bar{v}_t \}$ where $T \subset \mathcal{D}$. The attacker's goal is to promote the items in $T$, i.e., increase their estimated frequencies. Denoting by $\tilde{f}_b$ and $\tilde{f}_a$ the frequencies estimated by the LDP protocol before and after attack, MGA defines the frequency gain $\Delta \tilde{f}(\bar{v})$ as $\Delta \tilde{f}(\bar{v}) = \tilde{f}_a(\bar{v}) - \tilde{f}_b(\bar{v})$. Then, the overall Gain of the attack is: $\text{Gain} = \sum_{\bar{v} \in T} \Delta \tilde{f}(\bar{v})$. The attacker's goal is to maximize Gain.

\begin{algorithm}[!t]
\caption{Maximal Gain Attack on OLH}
\label{alg:MGA}
\begin{algorithmic}[1]
\Require $\mathcal{D}$, $T$, $\kappa$, $\mathcal{H}$
\Ensure $\langle H_{u}, x_{u}' \rangle$ for a poisoning user $u$
\State Initialize $max\_mp\_size$ as 0
\State Initialize current $\langle H_{u}, x_{u}' \rangle$ as null
\For{$\kappa$ number of iterations}
\State Randomly select a hash function $H$ from family $\mathcal{H}$
\State Initialize $counts$ as a list of integers, containing all zeros, with length $g$
\For{$\bar{v} \in T$}
\State $x \gets H(\bar{v})$
\State $counts[x] \gets counts[x] + 1$
\EndFor
\State $mp\_size \gets \max(counts)$
\State $mp\_size\_index \gets \argmax(counts)$
\If{$mp\_size > max\_mp\_size$}
\State Assign $\langle H_{u}, x_{u}' \rangle \gets \langle H, mp\_size\_index \rangle$  
\State Assign $max\_mp\_size \gets mp\_size$
\EndIf
\EndFor
\State \Return $\langle H_{u}, x_{u}' \rangle$
\end{algorithmic}
\end{algorithm}

Recall from the BIA that when a certain $\langle H_u, x_u' \rangle$ is reported to the server and we divide $\mathcal{D}$ into two $\mathcal{D}_1, \mathcal{D}_2$ such that $\mathcal{D}_1 = \{ v | v \in \mathcal{D}, H_u(v) = x'_u \}$ and $\mathcal{D}_2 = \mathcal{D} \setminus \mathcal{D}_1$, it holds that: $\text{Pr}[v_u \in \mathcal{D}_1] > \text{Pr}[v_u \in \mathcal{D}_2]$. Because of this, the strategy that maximizes Gain in OLH is to find the $\langle H_u, x_u' \rangle$ which maps as many of the items in $T$ to the same $x_u'$ as possible (from an attacker's perspective, the best scenario is when $\forall \bar{v}_i \in T$, $H_u(\bar{v}_i) = x_u'$). This enables $Sup(\bar{v}_i)$ to increase $\forall \bar{v}_i \in T$ when performing server-side estimation in OLH. Consequently, $\tilde{f}(\bar{v})$ is also increased, and the attack becomes more successful.

Following \cite{mga}, MGA for OLH can be implemented as shown in Algorithm \ref{alg:MGA}, i.e., malicious user $u$ determines their $\langle H_{u}, x_{u}' \rangle$ using Algorithm \ref{alg:MGA}. The inputs are the domain $\mathcal{D}$, target items $T$, iteration count parameter $\kappa$, and the hash function family $\mathcal{H}$. Following \cite{mga}, the value of $\kappa$ can be 1000. In each of the $\kappa$ total number of iterations, a hash function $H$ is drawn from $\mathcal{H}$. Then, the algorithm studies how many items from $\bar{v} \in T$ this $H$ hashes to each integer in $[0,g-1]$. These item counts are stored in a list called $counts$. The maximum count (denoted by $mp\_size$ in Alg.~\ref{alg:MGA}) and the index of the maximum count (denoted by $mp\_size\_index$) are found from $counts$. If the current $mp\_size$ exceeds the previously found maximum $max\_mp\_size$, then $\langle H_{u}, x_{u}' \rangle$ is updated using the current hash function $H$ and the current $mp\_size\_index$. At the end of $\kappa$ iterations performing the above, $\langle H_{u}, x_{u}' \rangle$ stores the hash function that hashes the maximum number of items from $T$ to $x_{u}'$. This $\langle H_{u}, x_{u}' \rangle$ is the tuple which maximizes Gain; therefore, it is produced as the output of Alg.~\ref{alg:MGA}.  


\vspace{-5pt}
\section{Exposing Hash-Induced Unfairness in LDP} \label{sec:Exposing}
\vspace{-3pt}

In this section, we empirically show that the impacts of BIA and MGA can differ significantly under the same $\varepsilon$ budget, due to variations in hash functions. First, we describe our experiment setup (datasets and metrics) in Section \ref{sec:ExperimentSetup}. Second, we describe the construction of subpopulations from $\mathcal{U}$ in Section \ref{sec:Subpops}. Third, we experimentally demonstrate the impacts of BIAs and MGAs on different subpopulations in Section \ref{sec:impacts}. 

\vspace{-4pt}
\subsection{Experiment Setup} \label{sec:ExperimentSetup}

\underline{Datasets.} We used three real-world datasets (Adult, BMS-POS, and Kosarak) and two synthetically generated datasets (Gaussian and Uniform) in our experiments. All implementations were done in Python. In all figures, we report average results from 10 experiments to account for LDP randomness. 

\textbf{Adult:} We used the Adult dataset from the UCI Machine Learning Repository\footnote{\url{https://archive.ics.uci.edu/dataset/2/adult}}, which contains demographic and employment-related information about individuals. We dropped records with missing data, yielding $|\mathcal{U}|$ = 45222, and then used the cross product of age and gender attributes as $\mathcal{D}$.

\textbf{BMS-POS:} The BMS-POS dataset contains sales data from a large electronics retailer, comprising 515596 transactions and 1657 distinct items sold\footnote{\url{https://github.com/cpearce/HARM/blob/master/datasets/BMS-POS.csv}}. For our experiments, we retained only the top 100 most frequently purchased items, hence $|\mathcal{D}|$ = 100. After this filtering, we had $|\mathcal{U}|$ = 400577 users remaining. 

\textbf{Kosarak:} The Kosarak dataset consists of click-stream data from a Hungarian online news portal\footnote{\url{https://www.philippe-fournier-viger.com/spmf/index.php?link=datasets.php}}. Since many URLs appeared infrequently (e.g., only once or twice), we pre-processed the dataset by selecting the 128 most frequently visited URLs and discarded the rest, hence $|\mathcal{D}|$ = 128. For users with multiple URLs remaining in their click stream sequence, we assigned the most frequently visited URL as their $v_u$.

\textbf{Gaussian:} We generated multiple synthetic Gaussian datasets by sampling data points from Gaussian distributions with mean $\mu$ = 50 and varying standard deviation $\sigma$ = 1, 5, 7, and 10. By default, we used $|\mathcal{D}|$ = 100 and $|\mathcal{U}|$ = 100000. The default $\sigma$ is $\sigma$ = 7, but we report results with varying $\sigma$ as well. 

\textbf{Uniform:} Lastly, we generated synthetic Uniform datasets where users' values are drawn from a Uniform distribution. We used $|\mathcal{U}|$ = 100000 and varied the domain size $|\mathcal{D}|$ between 100 and 2000.  

\underline{Metrics.} We used the Attack Success Rate (ASR) metric to measure the success of a BIA, and the ULoss metric to measure the success of a MGA. 

ASR captures the ratio of the population for which the attacker's predicted $v_u^p$ is correct. Higher ASR indicates lower privacy for users.
\begin{equation}
    \text{ASR} = \frac{\# \text{ of users } u \in \mathcal{U} \text{ such that } v_u = v_u^p}{|\mathcal{U}|}
\end{equation}

Since the goal of a MGA is to distort items' frequencies, the ULoss metric is defined to measure the differences between items' original and aggregated frequencies. The $\ell_1$ distance is used as the distance measure:
\begin{equation}
    \text{ULoss} = \sum\limits_{v \in \mathcal{D}} |\tilde{f}(v)-f(v)|
\end{equation}

\vspace{-6pt}
\subsection{Division of Users into Subpopulations} \label{sec:Subpops}

Since our goal is to expose unfairness in LDP caused by hash function behaviors, we divide the total user population $\mathcal{U}$ into subpopulations based on how the users' hash functions behave. More specifically, we construct three subpopulations: High-ENT, Low-PIS, and High-PIS, as explained below. 

\textbf{High-ENT (High Entropy)} consists of users whose hash functions $H_u$ distribute the values in $\mathcal{D}$ in a more uniform manner. To quantify uniformity, we use the notion of entropy. Higher entropy indicates a more uniform distribution of $\mathcal{D}$ into possible hash outputs $[0,g-1]$. We use the process outlined in Algorithm \ref{alg:entropy} to calculate the entropy of a hash function $H$. Initially, we set the counts of all $i \in [0,g-1]$ to zero. For each value $v$ in $\mathcal{D}$, we apply the hash function on $v$ to obtain $x \gets H(v)$, where $x \in [0,g-1]$. We then increment $count(x)$ by 1. After going through all $v \in \mathcal{D}$, we obtain the $count$s storing how many elements in $\mathcal{D}$ hash to each integer $i$. We convert $count(i)$ to $prob(i)$ (lines 6-7), and compute the entropy of the hash function, denoted by $\mathcal{E}_\text{comp}$, on line 8.

\begin{algorithm}[!t]
\caption{Calculating the Entropy of a Hash Function}
\label{alg:entropy}
\begin{algorithmic}[1]
\Require Hash function $H$, domain $\mathcal{D}$
\Ensure Computed entropy of this hash function $\mathcal{E}_\text{comp}$
\For{$i = 0$ to $g-1$}
\State Initialize $count(i) = 0$
\EndFor
\For{$v \in \mathcal{D}$}
\State $x \gets H(v)$
\State $count(x) \gets count(x) + 1$
\EndFor
\For{$i = 0$ to $g-1$}
\State $prob(i) \gets \frac{count(i)}{|\mathcal{D}|}$
\EndFor
\State $\mathcal{E}_\text{comp} \gets - \sum\limits_{i=0}^{g-1} prob(i) \times \text{log}(prob(i))$
\State \Return $\mathcal{E}_\text{comp}$
\end{algorithmic}
\end{algorithm}

Given the total user population $\mathcal{U}$, the High-ENT subpopulation is constructed as follows. For each user $u \in \mathcal{U}$, we feed the user's hash function $H_u$ into Algorithm \ref{alg:entropy} to obtain the entropy for this user, $\mathcal{E}_\text{comp}^{u}$. After obtaining the entropies for all users $\mathcal{E}_\text{comp}^{u_1}$, $\mathcal{E}_\text{comp}^{u_2}$, ..., $\mathcal{E}_\text{comp}^{u_{|\mathcal{U}|}}$ we sort them in descending order. We select the top 10\% of the users with the highest entropies to constitute the set of users in High-ENT. 

\textbf{Low-PIS (Low Preimage Size).} Recall from the OLH protocol that we have $x_u = H_u(v_u)$ for user $u$. The preimage set of $H_u$, which we denote by $P_u$, can be defined as: $P_u = \{v | v \in \mathcal{D}, H_u(v) = x_u \}$. In other words, $P_u$ is the set of values in the domain that hash to the same outcome as $v_u$. Low-PIS consists of users whose hash functions yield low $P_u$. Given the total user population $\mathcal{U}$, the Low-PIS subpopulation is constructed as follows. For each user $u \in \mathcal{U}$, we calculate the $P_u$ for this user. After obtaining all preimage set sizes $P_{u_1}$, $P_{u_2}$, ..., $P_{u_{|\mathcal{U}|}}$, we sort them in descending order. We select the bottom 10\% of the users with the lowest preimage set sizes to constitute the set of users in Low-PIS. 

\textbf{High-PIS (High Preimage Size).} We use the same preimage set sizes $P_u$ as in Low-PIS, but this time, we focus on the opposite end, i.e., users whose hash functions yield high $P_u$. After obtaining $P_{u_1}$, $P_{u_2}$, ..., $P_{u_{|\mathcal{U}|}}$ and sorting them in descending order, we select the top 10\% of the users with the highest preimage set sizes to constitute the set of users in High-PIS. 

\underline{Summary.} We summarize the intuition behind High-ENT, Low-PIS, and High-PIS as follows. Users in High-ENT are those users in $\mathcal{U}$ whose hash functions yield the most uniform-like output distributions, i.e., values in $\mathcal{D}$ are as evenly distributed as possible over the output space $[0,g-1]$. On the other hand, Low-PIS and High-PIS both correspond to cases where hash function non-uniformity is high, i.e., certain outputs are more likely than others. Yet, they represent two opposite ends: In High-PIS, non-uniformity is high \textit{and} there are many hash collisions with $x_u = H_u(v_u)$. In Low-PIS, non-uniformity is high \textit{but} as opposed to High-PIS, there are very few hash collisions with $x_u$. 


\subsection{Impacts of BIA and MGA on Different Subpopulations} \label{sec:impacts}

\begin{figure}[!t]
  \centering
  \begin{minipage}[b]{0.24\textwidth}
    \includegraphics[width=\textwidth]{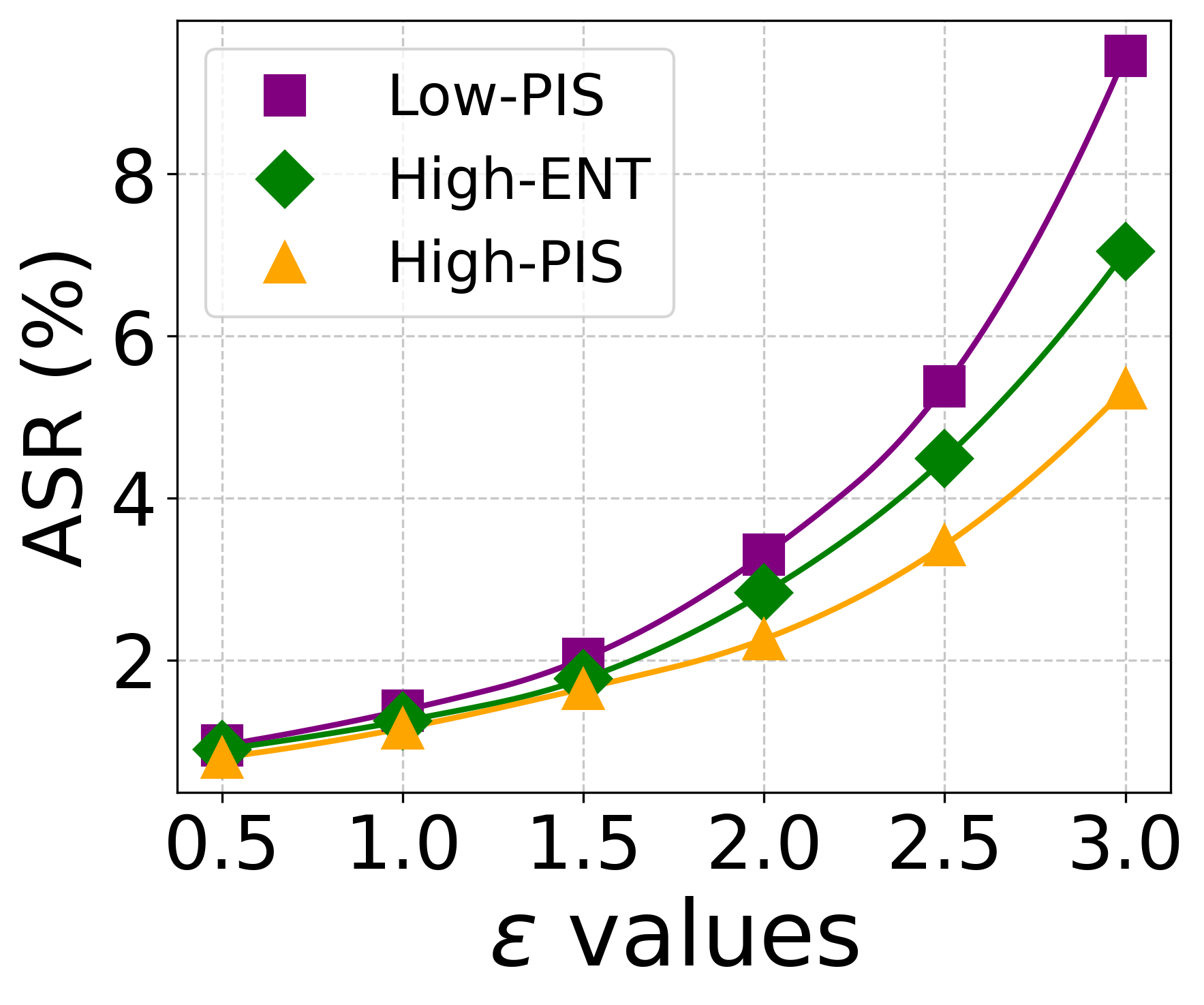}
  \end{minipage}
  \hfill
  \begin{minipage}[b]{0.24\textwidth}
    \includegraphics[width=\textwidth]{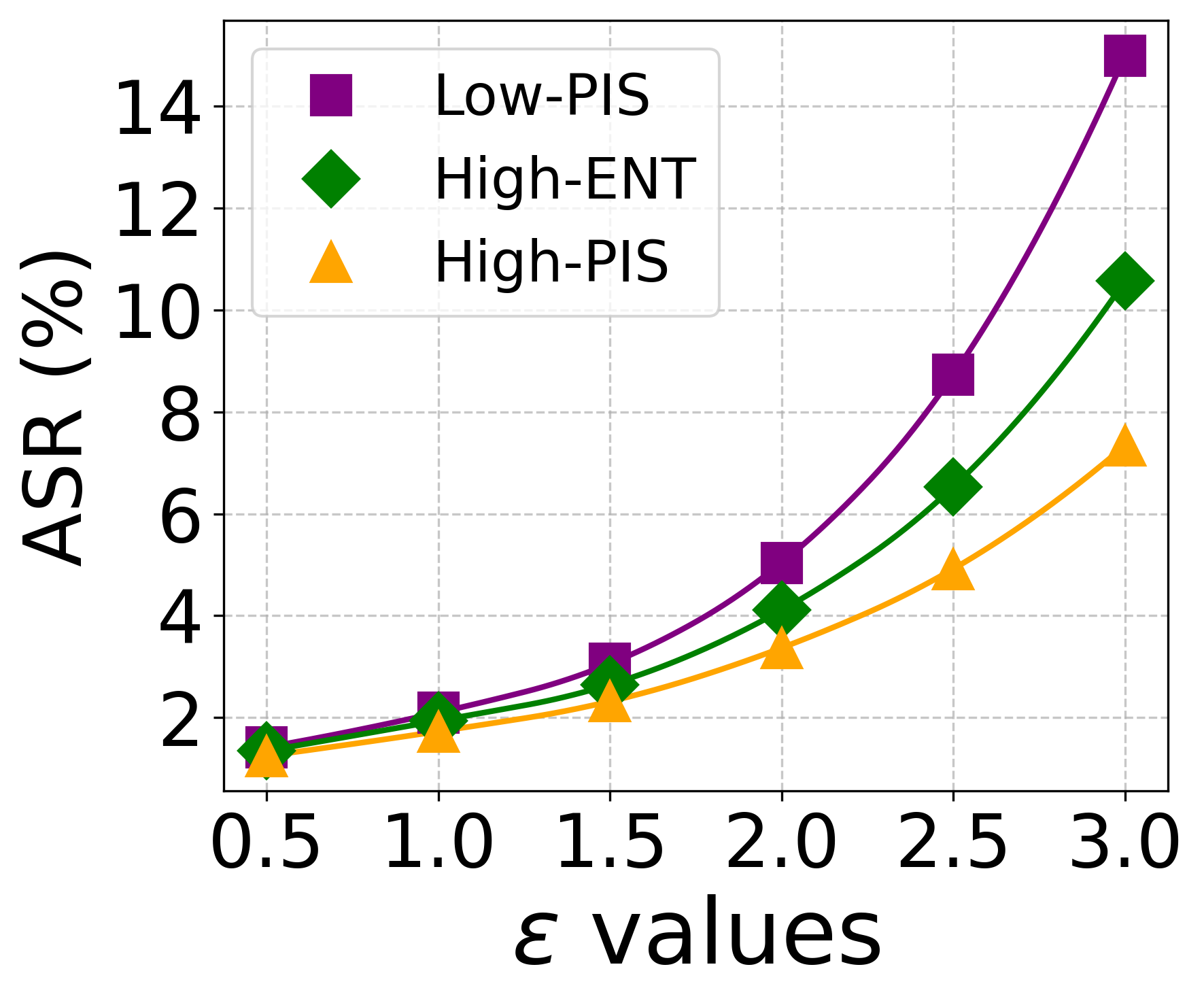}
  \end{minipage}
  \hfill
  \begin{minipage}[b]{0.24\textwidth}
    \includegraphics[width=\textwidth]{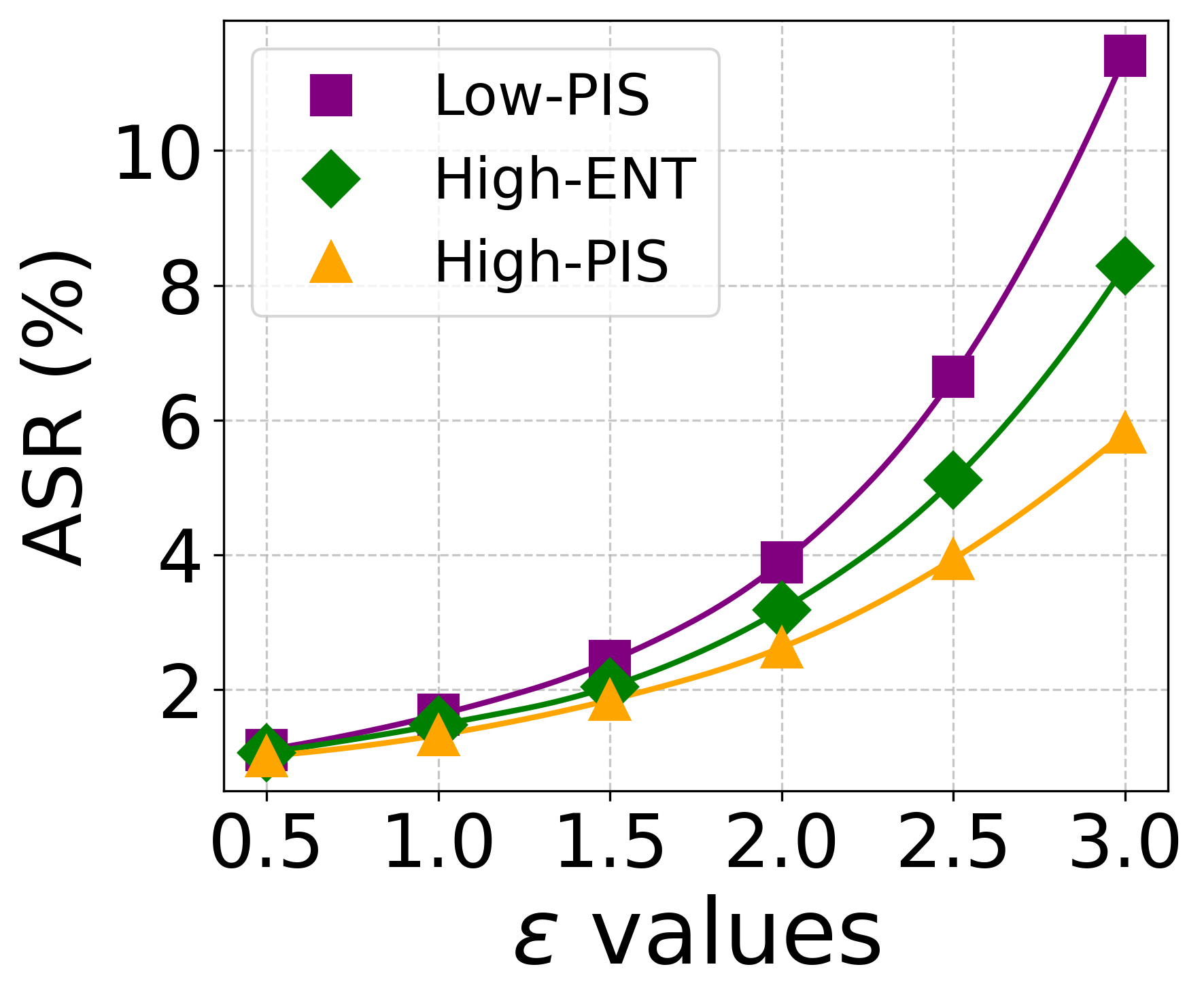}
  \end{minipage}
  \hfill
  \begin{minipage}[b]{0.24\textwidth}
    \includegraphics[width=\textwidth]{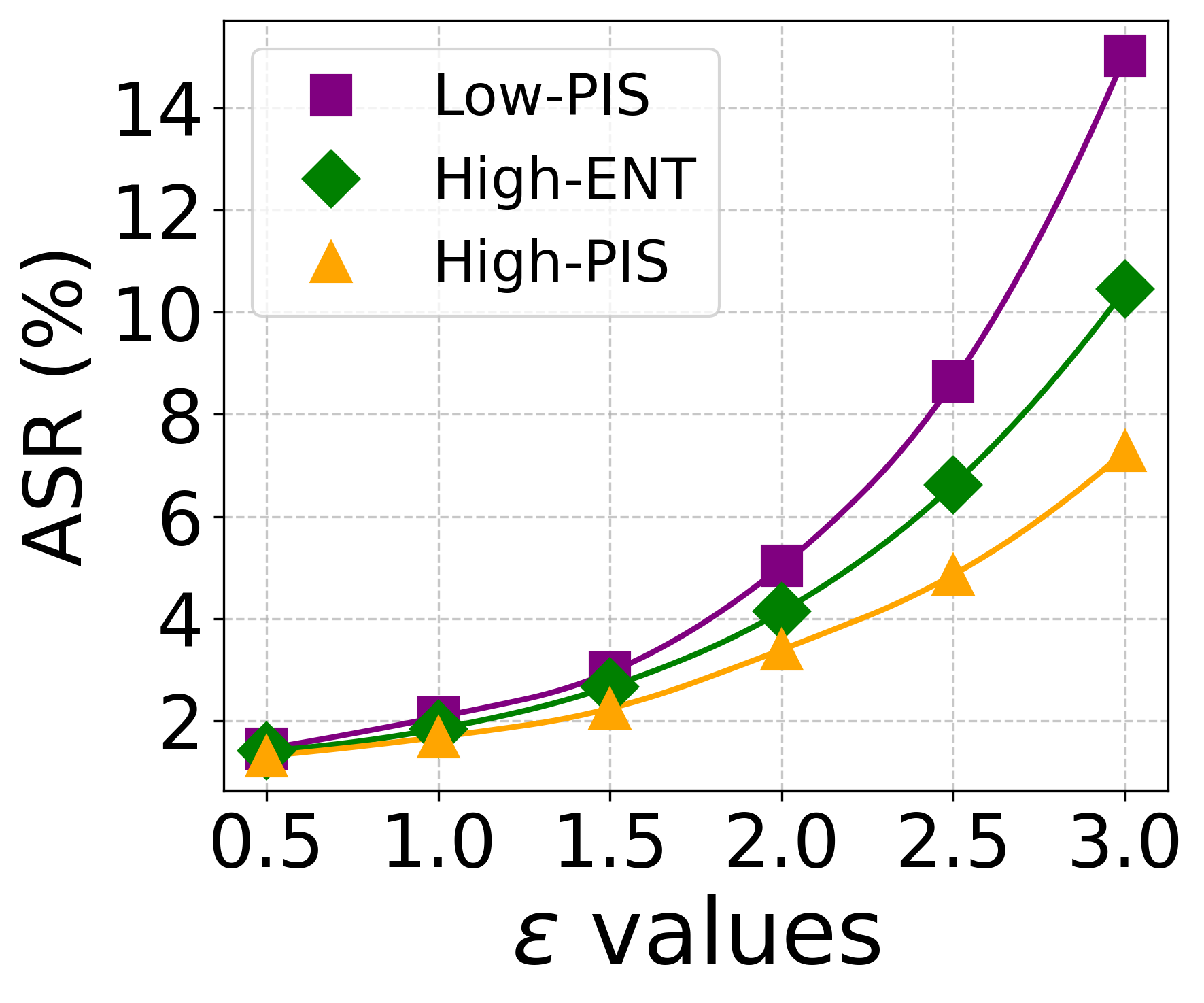}
  \end{minipage}
  \vspace{-6pt}
  \caption{BIA results on Adult, BMS-POS, Kosarak, and Gaussian datasets.}
  \vspace{-2pt}
  \label{fig:bia}
\end{figure}

We execute BIAs on the three subpopulations (High-ENT, Low-PIS, and High-PIS) and measure their ASRs separately. Results with varying $\varepsilon$ are shown in Figure \ref{fig:bia}. The results show that significant differences in ASR can be observed for different subpopulations, especially when $\varepsilon$ is increased. When $\varepsilon$ is low, all subpopulations' ASRs remain low, therefore the differences between them are not too visible in the graphs (though they exist). As $\varepsilon \geq 2$, the differences become more visible. When $\varepsilon$ = 3, ASRs for Low-PIS are typically 2x higher than High-PIS and 1.5x higher than High-ENT. These validate that the behaviors of hash functions indeed have a substantial impact on ASRs, and users in the Low-PIS population can be twice as vulnerable to a BIA compared to High-PIS users, despite using the same LDP protocol and $\varepsilon$ budget. 

It is intuitive that Low-PIS users have higher vulnerability than others. By definition of Low-PIS, users in this subpopulation use hash functions with small preimage sets $P_u$. From Algorithm \ref{alg:BIA}, we observe that if $P_u$ is small, then lines 5-6 of the algorithm will increment the scores of fewer different values, since the condition on line 5 will be satisfied by only those $v \in P_u$. Hence, the argmax on line 7 will recover the correct prediction. However, for High-PIS users, since $P_u$ is large, lines 5-6 of Algorithm \ref{alg:BIA} will increment the scores of many different values, and consequently, the argmax on line 7 will necessitate random tie-breaking or pick an incorrect prediction. High-ENT's ASR falls between these two subpopulations, because $P_u$ in High-ENT is neither as small as Low-PIS, nor as large as High-PIS. For example, $P_u$ in High-ENT can be close to $\frac{|\mathcal{D}|}{g}$, whereas it is larger in High-PIS and smaller in Low-PIS. Making a correct prediction from a larger set is more difficult, hence lower ASR in High-PIS. 

\begin{figure}[!t]
  \centering
  \begin{minipage}[b]{0.24\textwidth}
    \includegraphics[width=\textwidth]{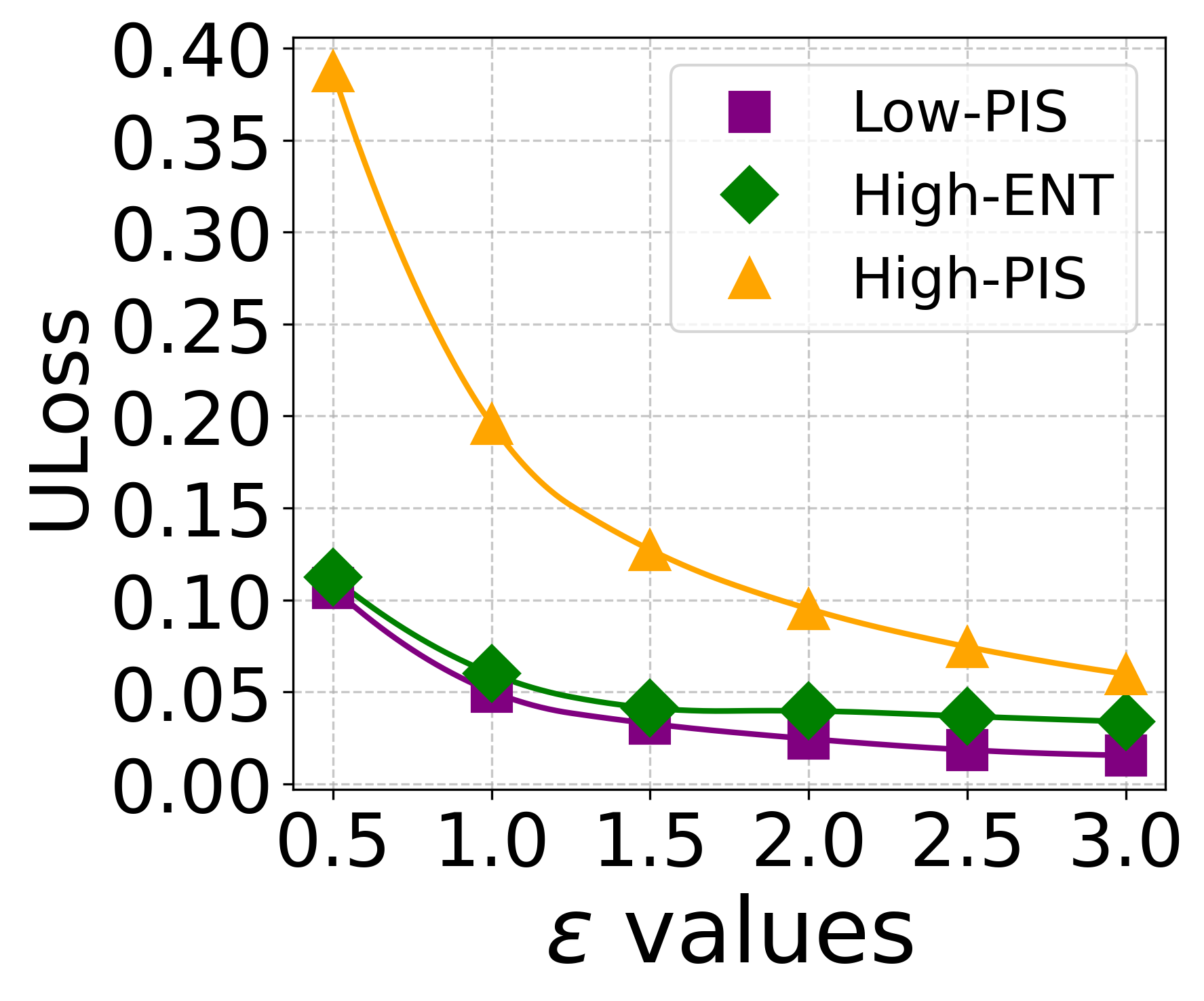}
  \end{minipage}
  \hfill
  \begin{minipage}[b]{0.24\textwidth}
    \includegraphics[width=\textwidth]{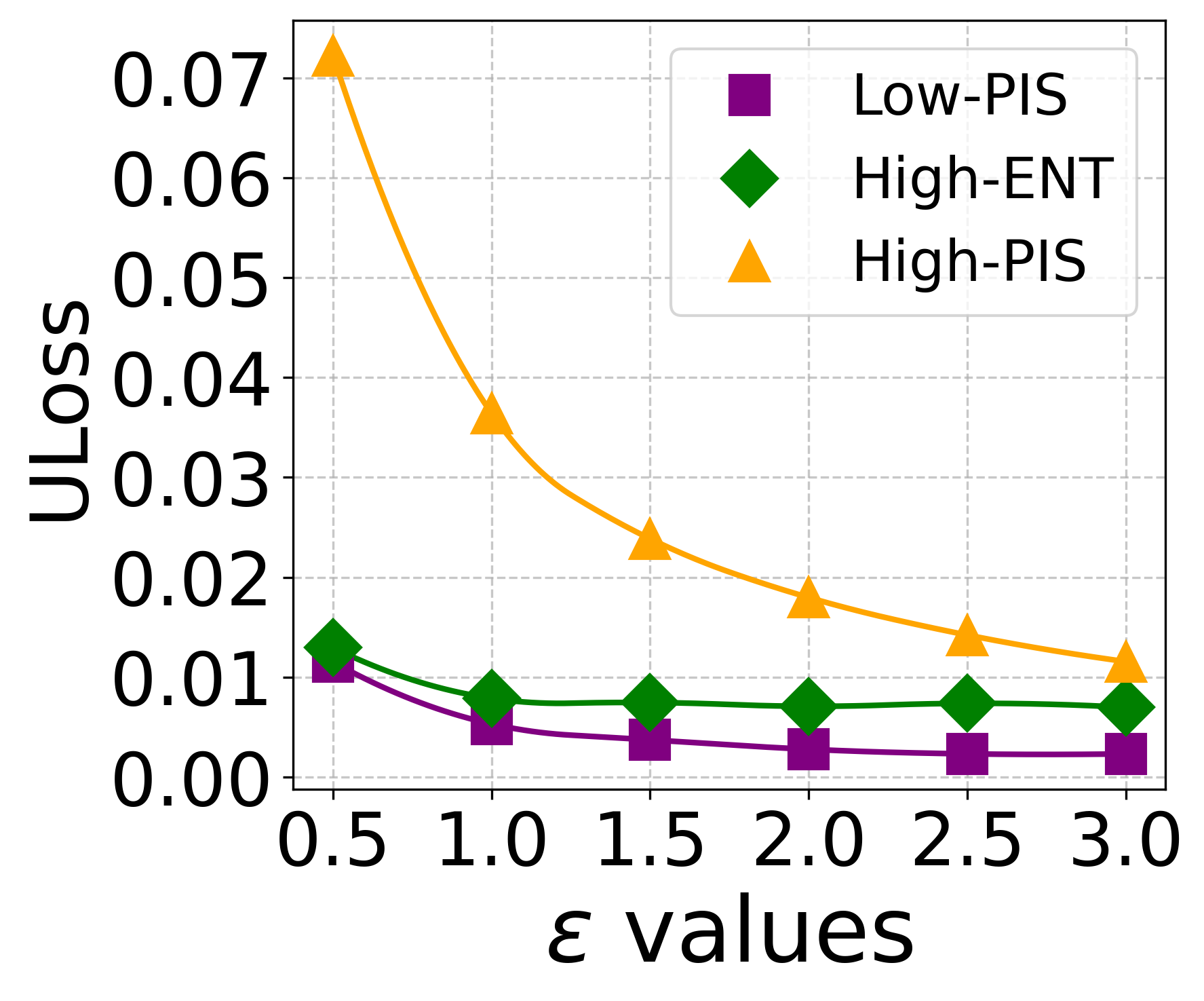}
  \end{minipage}
  \hfill
  \begin{minipage}[b]{0.24\textwidth}
    \includegraphics[width=\textwidth]{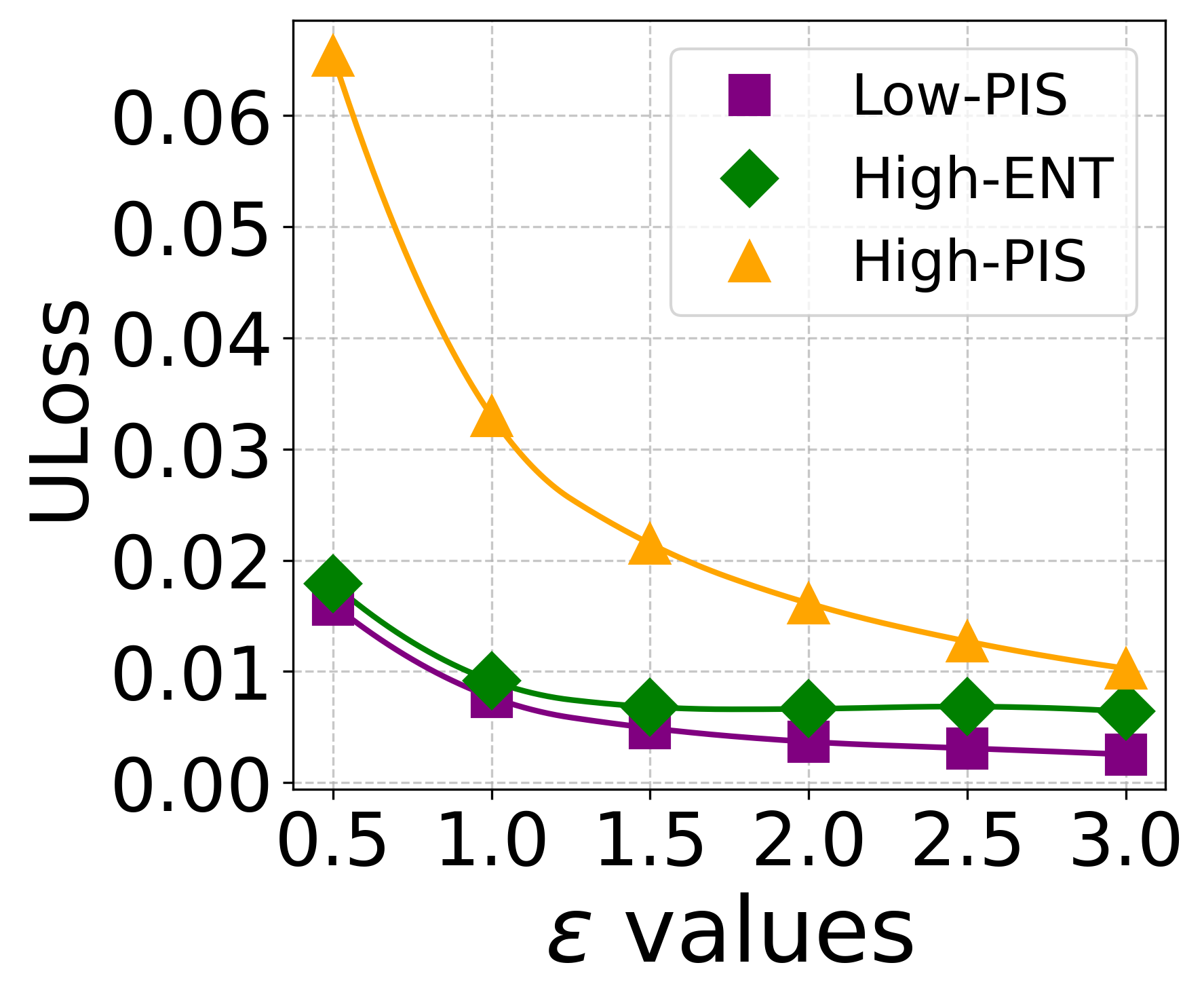}
  \end{minipage}
  \hfill
  \begin{minipage}[b]{0.24\textwidth}
    \includegraphics[width=\textwidth]{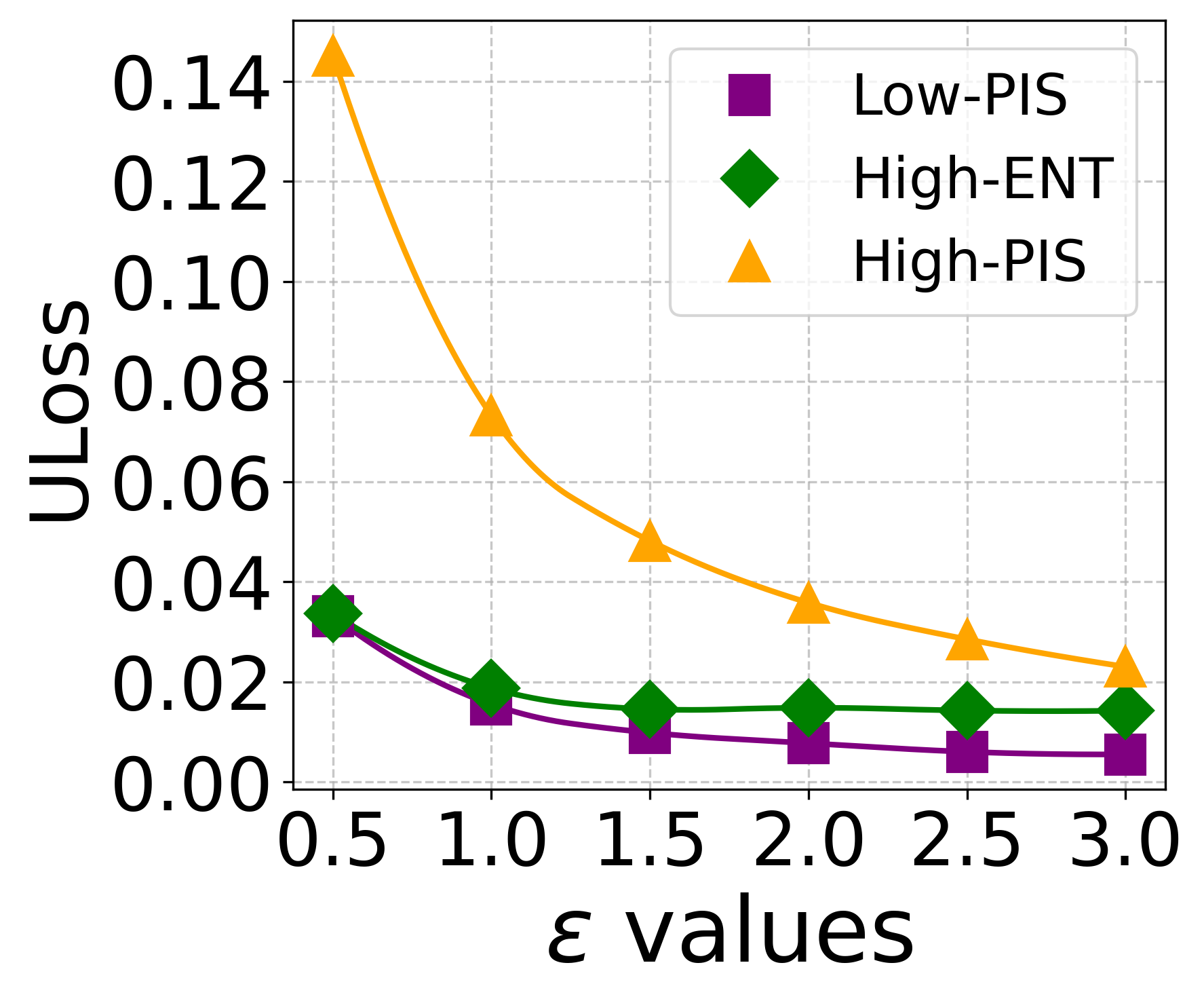}
  \end{minipage}
  \vspace{-6pt}
  \caption{MGA results on Adult, BMS-POS, Kosarak, and Gaussian datasets.}
  \label{fig:mga}
  \vspace{-4pt}
\end{figure}

Next, we execute MGAs on the three subpopulations and measure their ULoss separately. To make MGA untargeted, we use $T$ = $\mathcal{D}$. Results of this experiment with varying $\varepsilon$ are shown in Figure \ref{fig:mga}. According to the results, again there are significant differences in ULoss for different subpopulations. ULoss of High-PIS is especially much higher than Low-PIS and High-ENT. Conversely, Low-PIS exhibits the lowest ULoss, while the ULoss of High-ENT falls between Low-PIS and High-PIS. Consequently, we find that if the malicious user in a MGA is a High-PIS user, they can achieve a much more successful attack than a Low-PIS or High-ENT user, e.g., the attack can increase ULoss by 3-4 times more when $\varepsilon \leq 1$. 

The reason behind these results is that the hash function and the resulting $P_u$ are important in MGA as well. Consider lines 6-10 of Algorithm \ref{alg:MGA}. When $P_u$ is large, it becomes more likely that many $\bar{v} \in T$ are also members of $P_u$. Consequently, the final $\langle H_{u}, x_{u}' \rangle$ returned by Algorithm \ref{alg:MGA} affects the resulting frequencies $\tilde{f}(v)$ more. This causes a larger discrepancy between $\tilde{f}(v)$ and $f(v)$, resulting in higher ULoss. This trend is quite amplified in High-PIS; therefore, High-PIS has much higher ULoss than High-ENT and Low-PIS. 

\vspace{-4pt}
\section{F-OLH: Mitigating Hash-Induced Unfairness} \label{sec:Mitigating}
\vspace{-2pt}

Having shown that users' membership in Low-PIS or High-PIS yields substantial differences in terms of BIA and MGA, we now shift our direction towards how these disparate impacts (unfairness) can be mitigated. To address this problem, we propose a fair variant of OLH, called Fair-OLH (F-OLH). 

\vspace{-4pt}
\subsection{Intuition Behind F-OLH}

\begin{figure}[!t]
\centering
\begin{minipage}[b]{0.325\textwidth}
    \includegraphics[width=\textwidth]{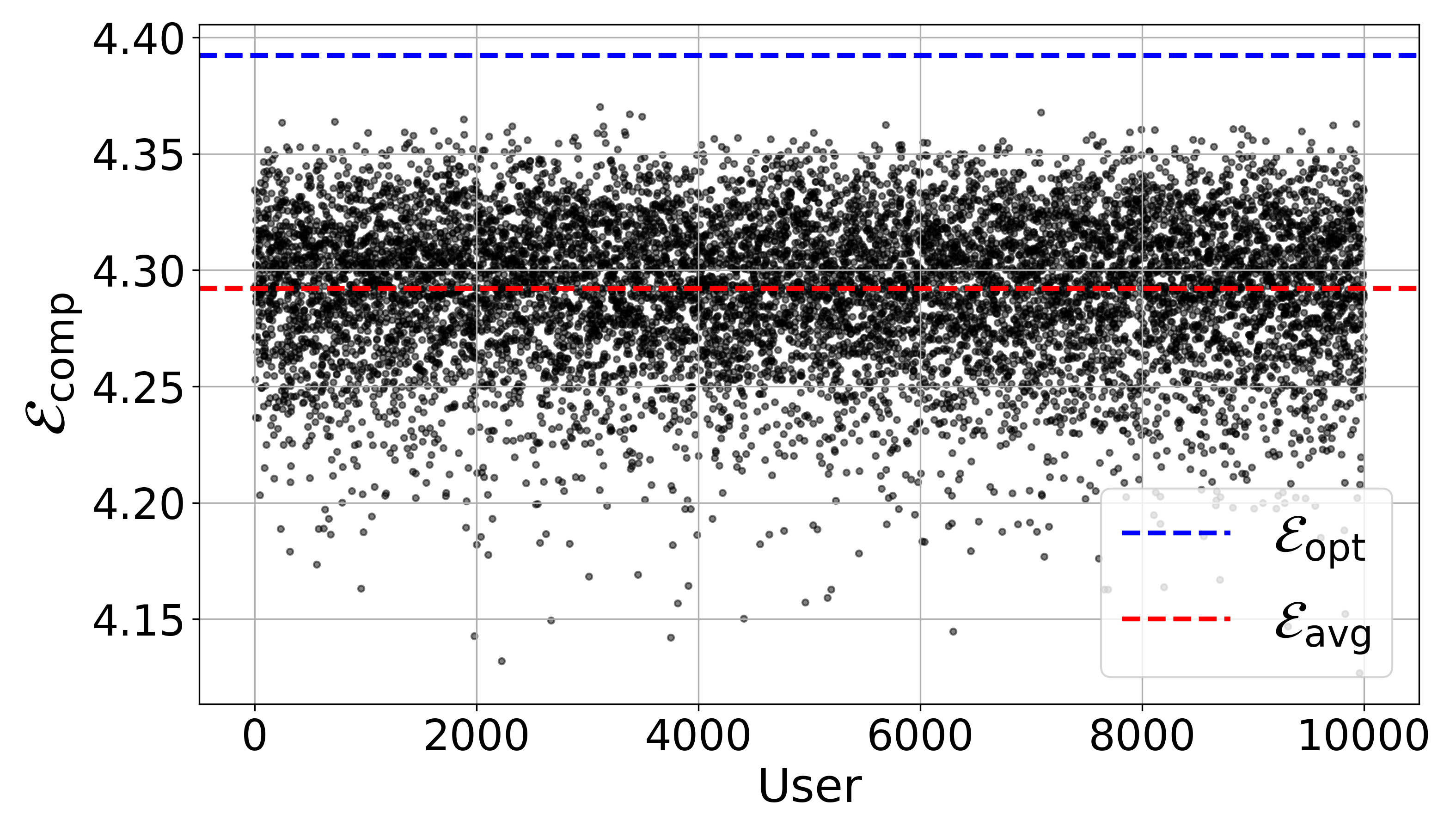}
  \end{minipage}
  \hfill
  \begin{minipage}[b]{0.325\textwidth}
    \includegraphics[width=\textwidth]{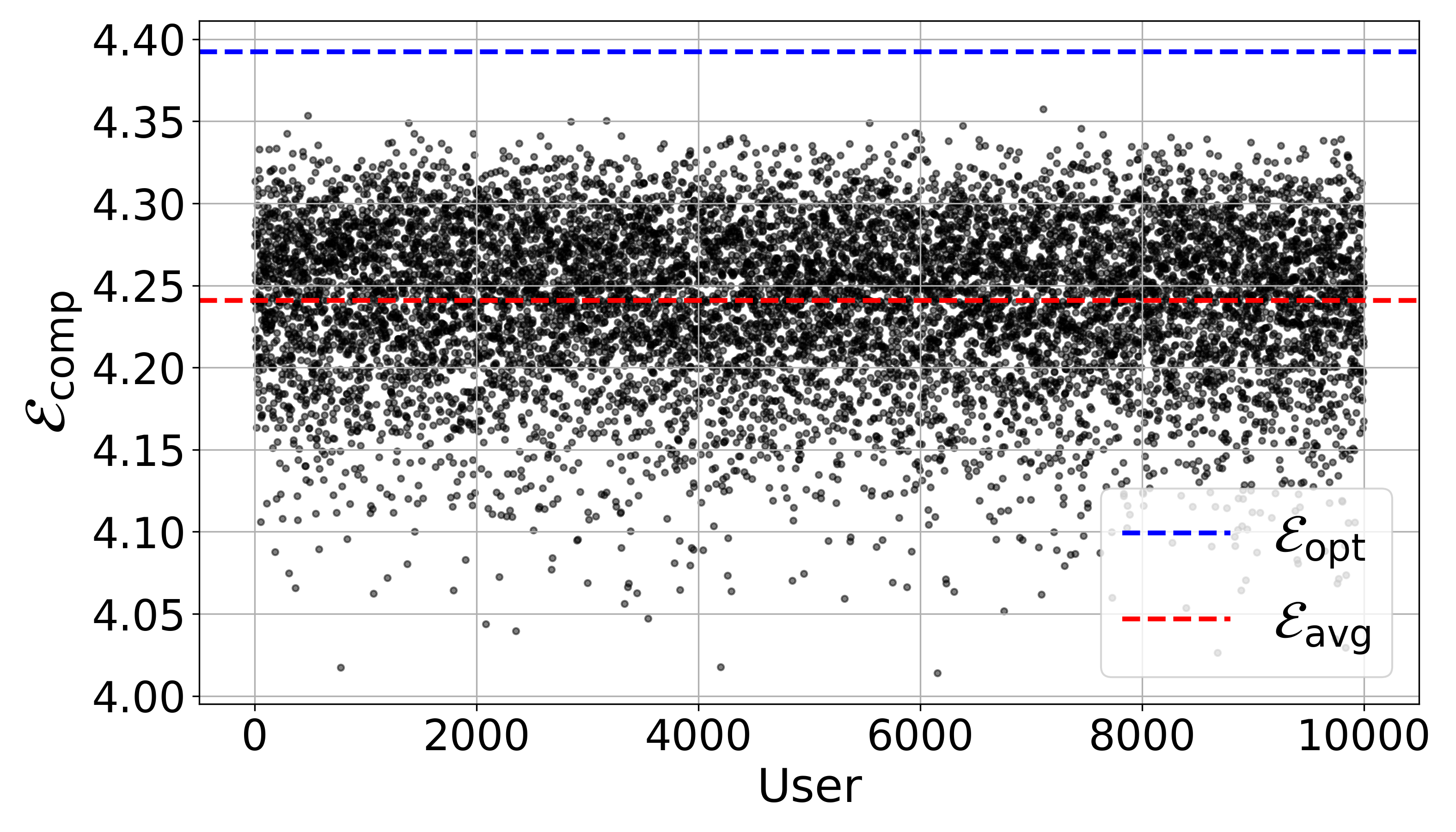}
  \end{minipage}
  \hfill
  \begin{minipage}[b]{0.325\textwidth}
    \includegraphics[width=\textwidth]{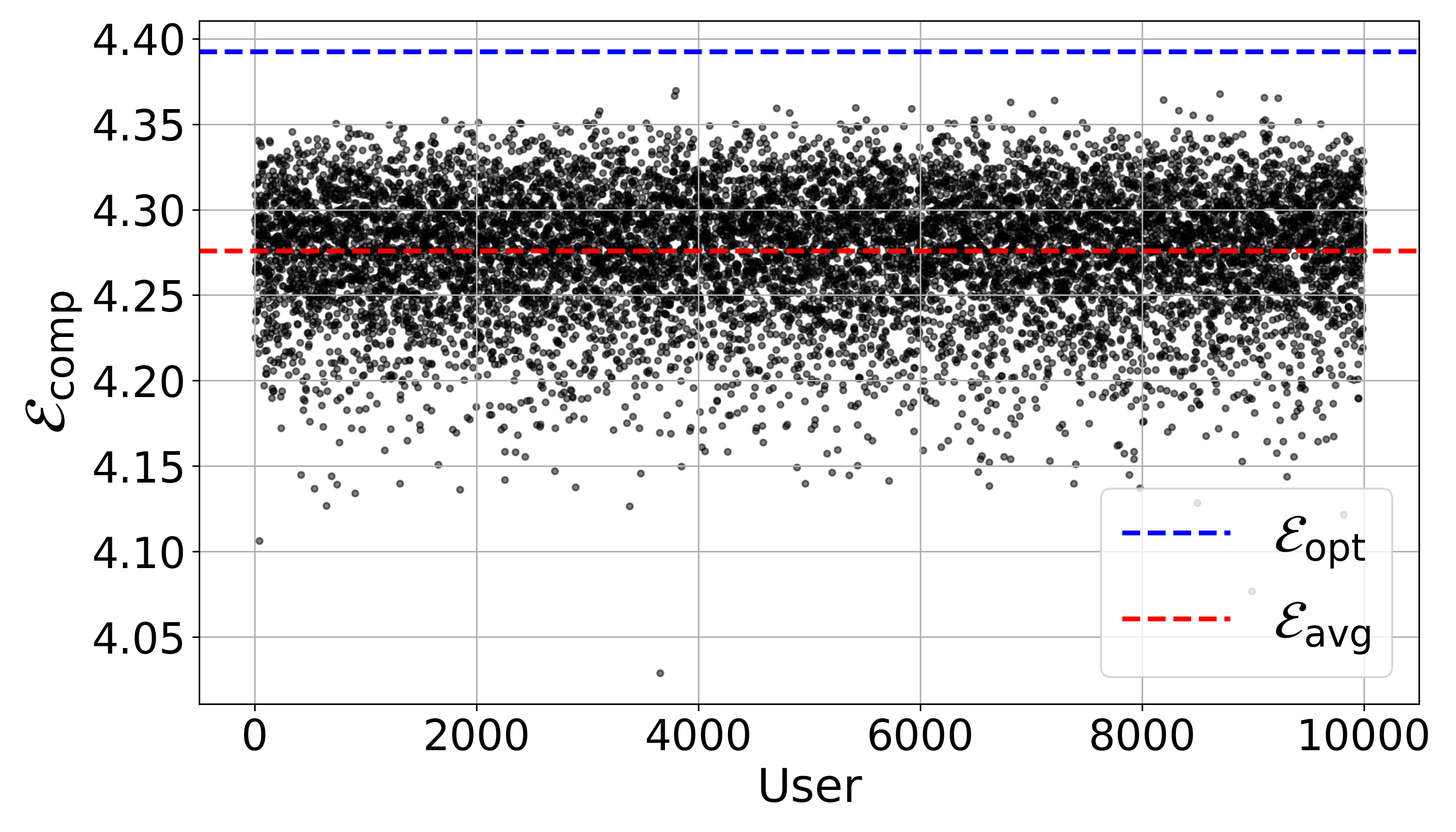}
  \end{minipage} \\
  \begin{minipage}[b]{0.325\textwidth}
    \includegraphics[width=\textwidth]{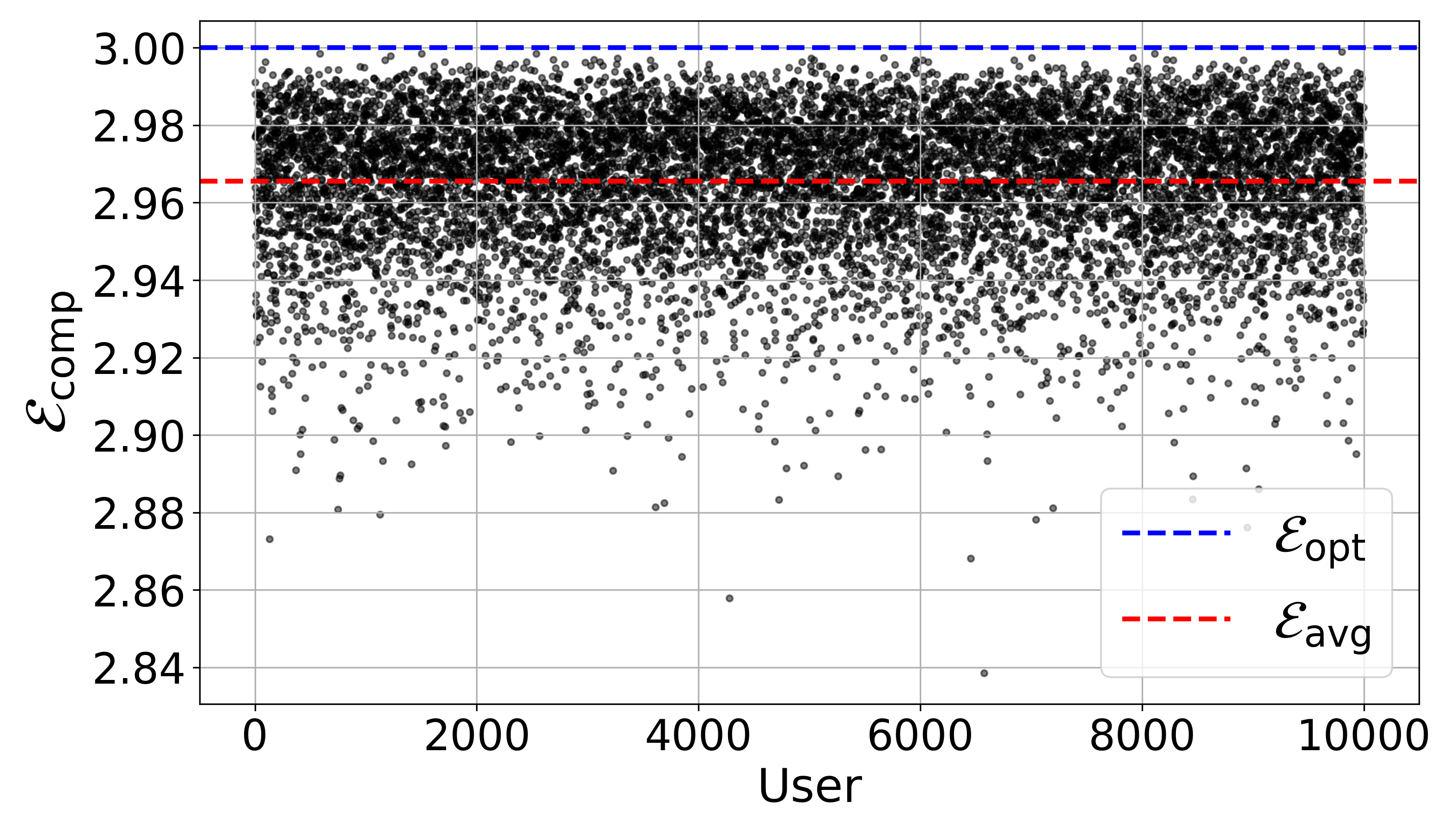}
  \end{minipage} 
  \hfill
  \begin{minipage}[b]{0.325\textwidth}
    \includegraphics[width=\textwidth]{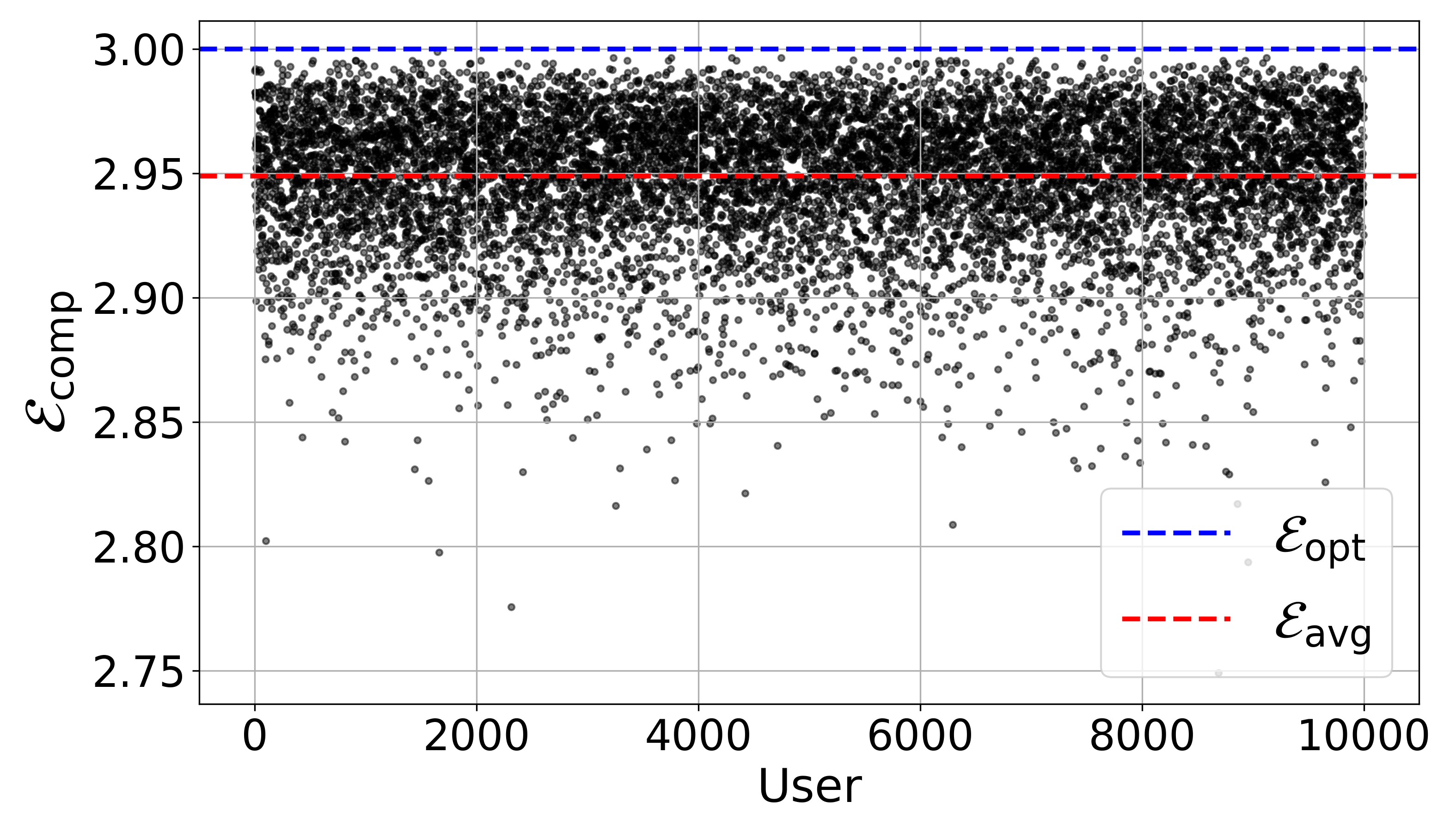}
  \end{minipage}
  \hfill
  \begin{minipage}[b]{0.325\textwidth}
    \includegraphics[width=\textwidth]{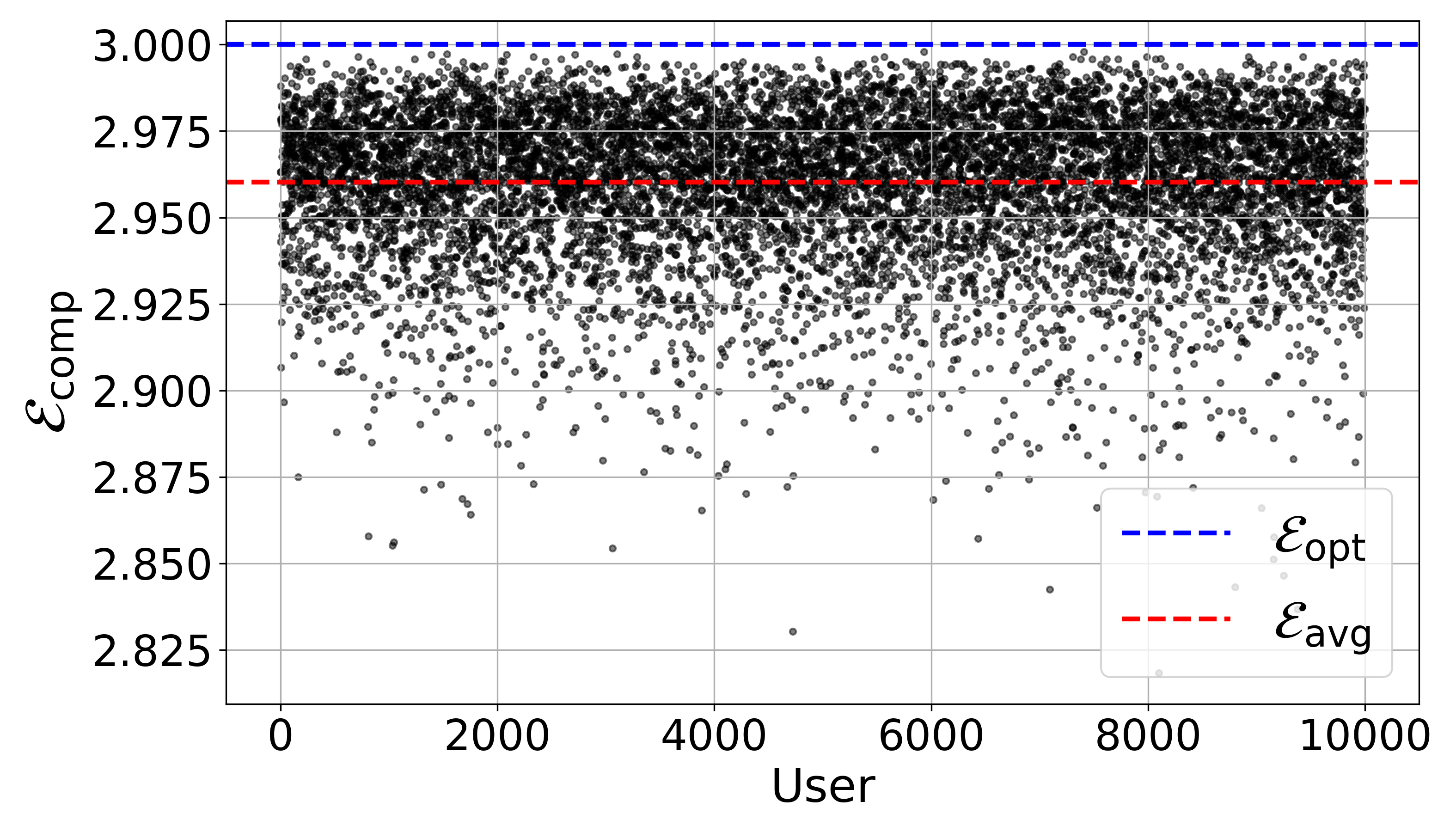}
  \end{minipage}
  \vspace{-6pt}
    \caption{Entropy of individual users' hash functions compared to $\mathcal{E}_\text{avg}$ and $\mathcal{E}_\text{opt}$. First row is with $\varepsilon$ = 3 and second row is with $\varepsilon$ = 2. Datasets from left to right: Adult, BMS-POS, Kosarak.}
  \label{fig:entropies}
  \vspace{-2pt}
\end{figure}

Considering that the unfairness is caused by the hash functions of different users behaving differently, we ask the following questions: How differently do different users' hash functions behave in practice? Is it possible to enforce that all users' hash functions behave similarly in terms of entropy? To illustrate the answers to these questions, we provide the results shown in Figure \ref{fig:entropies}. We sampled 10000 users from the corresponding user populations and computed the entropies of their hash functions $\mathcal{E}_\text{comp}$ using Algorithm \ref{alg:entropy}. Each user's $\mathcal{E}_\text{comp}$ is denoted by a black dot in Figure \ref{fig:entropies}. Then, we computed the average entropy of all users, denoted by $\mathcal{E}_\text{avg}$ and drawn using a dashed red line in Figure \ref{fig:entropies}. Finally, we computed the entropy of a hypothetical ideal hash function with maximum entropy, i.e., $\mathcal{D}$ is distributed evenly into outputs $[0, g-1]$. This is denoted by $\mathcal{E}_\text{opt}$ and drawn using a dashed blue line in Figure \ref{fig:entropies}. 

We observe from Figure \ref{fig:entropies} that when $\varepsilon$ = 2, some users' $\mathcal{E}_\text{comp}$ can be close to $\mathcal{E}_\text{opt}$. When $\varepsilon$ = 3, $\mathcal{E}_\text{comp}$ and $\mathcal{E}_\text{opt}$ are farther from one another because when $\varepsilon$ is increased, the $g$ parameter of the OLH protocol, which is tied to $\varepsilon$, also increases. Thus, the output space $[0,g-1]$ of the hash function enlarges. Consequently, it becomes more difficult for a hash function to achieve perfect uniformity in a larger output space. We also observe from Figure \ref{fig:entropies} that many users' hash functions behave similarly to $\mathcal{E}_\text{avg}$. Some have higher entropy and some have lower entropy than $\mathcal{E}_\text{avg}$, but they are usually not far from $\mathcal{E}_\text{avg}$. However, there also exist a few users whose $\mathcal{E}_\text{comp}$ is significantly low. These are the users who are typically members of the Low-PIS and High-PIS subpopulations, and therefore, they are the users with the highest amount of hash-induced unfairness. Therefore, our intuition is to design a solution which ensures $\mathcal{E}_\text{comp}$ is not too low. Indeed, our proposed F-OLH protocol stems from this principle. 


\subsection{Formalization of F-OLH} 

To mitigate hash-induced unfairness in LDP, we propose the Fair-OLH (F-OLH) protocol. The main idea of F-OLH is to enforce each user to select a hash function that ensures $\mathcal{E}_\text{comp}$ is not too low. For user $u$ with a certain hash function $H_u$, we define the fairness ratio $\rho_u$ as:
\begin{equation}
    \rho_u = \frac{\mathcal{E}_{\text{opt}}}{\mathcal{E}_{\text{comp}}}
\end{equation}
If $\rho_u$ is greater than a system-wide threshold $\rho$, F-OLH enforces the user to select another hash function. If $\rho_u$ with the newly selected hash function is again greater than $\rho$, another hash function is selected. This selection process keeps iterating until a suitable hash function that satisfies the $\rho$ threshold is found. Afterwards, the user executes the OLH protocol with the suitable hash function that was found.

\begin{algorithm}[!t]
\caption{Fair-OLH (F-OLH)}
\label{alg:FOLH}
\begin{algorithmic}[1]
\Require Domain $\mathcal{D}$, true value $v_u$, budget $\varepsilon$, hash function family $\mathcal{H}$, fairness ratio $\rho$
\Ensure Result of user-side perturbation $\langle H_u, x_u' \rangle$
\State $g \gets e^{\varepsilon} + 1$
\State $\mathcal{E}_\text{opt} \gets - \sum\limits_{i=0}^{g-1} \frac{1}{g} \times \text{log}(\frac{1}{g})$
\State $H_u \gets$ Draw a hash function from $\mathcal{H}$
\State $\mathcal{E}_{\text{comp}} \gets$ Call Algorithm \ref{alg:entropy} with $H_u$ and $\mathcal{D}$
\While{$\frac{\mathcal{E}_\text{opt}}{\mathcal{E}_{\text{comp}}} > \rho$}
\State $H_u \gets$ Draw a hash function from $\mathcal{H}$
\State $\mathcal{E}_{\text{comp}} \gets$ Call Algorithm \ref{alg:entropy} with $H_u$ and $\mathcal{D}$
\EndWhile
\State $x_u \gets H_u(v_u)$
\State Perturb $x_u$ to $x'_u \in [0,g-1]$ such that: 
$\Pr[x_u' = i] =
\begin{cases}
\frac{e^{\varepsilon}}{e^{\varepsilon} + g - 1}, & \text{if } x_u = i, \\
\frac{1}{e^{\varepsilon} + g - 1}, & \text{otherwise}
\end{cases}
$
\State Send $\langle H_u, x_u' \rangle$ to the server
\end{algorithmic}
\end{algorithm}

The pseudocode of F-OLH is given in Algorithm \ref{alg:FOLH}. The calculation of $g$ is identical to OLH. Then, $\mathcal{E}_\text{opt}$ is computed assuming an optimal hash function which uniformly distributes $\mathcal{D}$ to $[0,g-1]$. Note that in such a hash function, the probability of each hash outcome is $\frac{1}{g}$. Then, between lines 3-7, the user searches for a suitable $H_u$ which satisfies the $\frac{\mathcal{E}_\text{opt}}{\mathcal{E}_{\text{comp}}} \leq \rho$ condition by drawing hash functions from $\mathcal{H}$ and discarding those which do not satisfy the condition. Once a suitable hash function $H_u$ is found, the user hashes their true value $v_u$ using $H_u$ to obtain $x_u$. The perturbation of $x_u$ to $x'_u$ is identical to OLH. Finally, the user sends $\langle H_u, x_u' \rangle$ to the server. We note that this algorithm describes the user-side execution of F-OLH. On the server side, the server collects $\langle H_u, x_u' \rangle$ from all users $u \in \mathcal{P}$. The server's aggregation and estimation process remains identical to OLH, which was explained at the end of Section \ref{sec:LDP}. 

From an implementation standpoint, current implementations of the OLH protocol typically use a Python library like \textsc{xxhash} for hashing. In these implementations, drawing $H_u$ from $\mathcal{H}$ means changing the hash seed used in \textsc{xxhash}, since each different seed yields a different $H_u$. In F-OLH, we use a similar strategy to implement iterative drawing of different $H_u$ from $\mathcal{H}$, i.e., we iterate over different seeds in \textsc{xxhash} until Algorithm \ref{alg:FOLH} finds a suitable $H_u$. 

\subsection{Results of F-OLH on BIA and MGA}

\begin{figure}[t]
  \centering
  \begin{minipage}[b]{0.24\textwidth}
    \includegraphics[width=\textwidth]{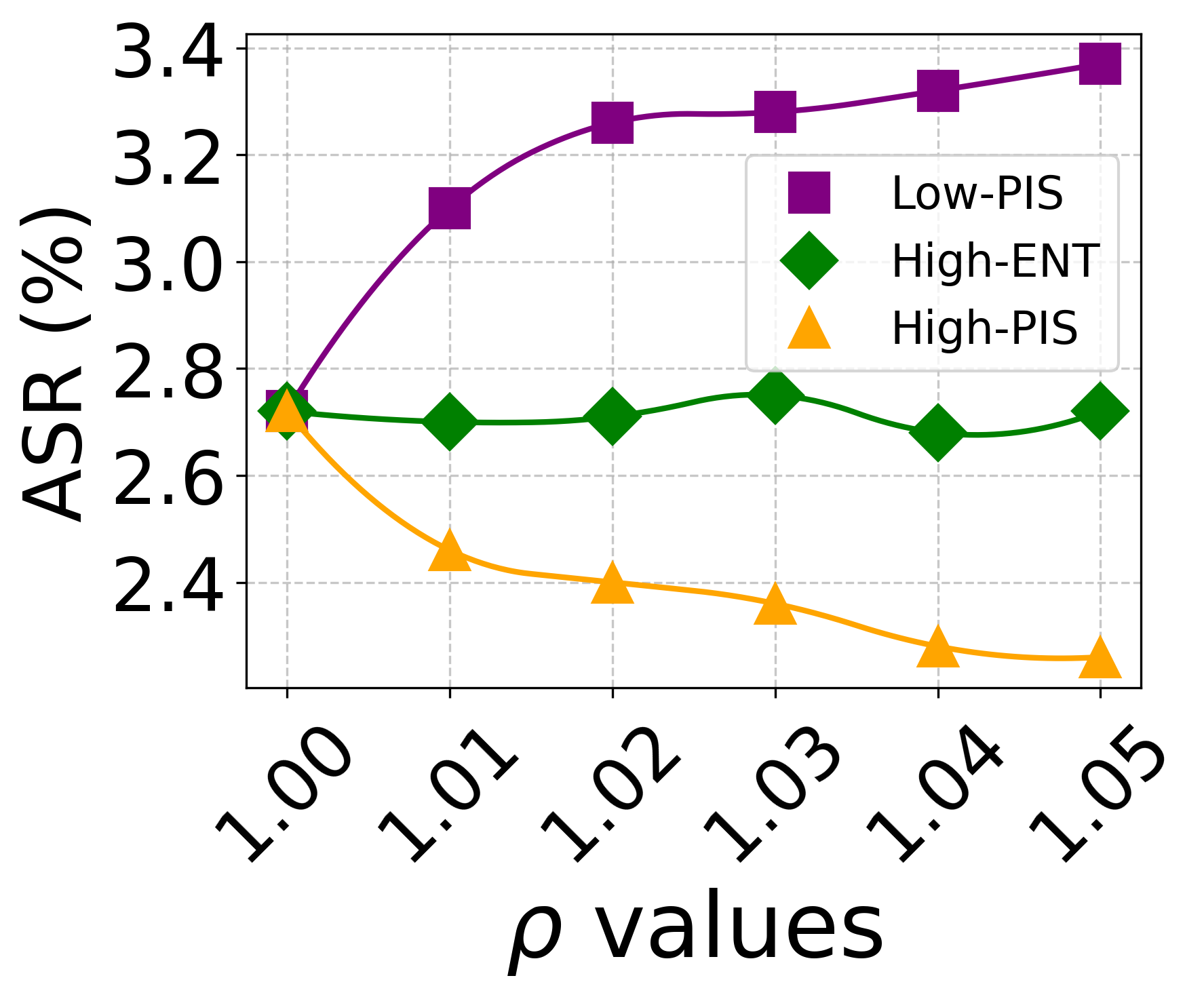}
  \end{minipage}
  \hfill
  \begin{minipage}[b]{0.24\textwidth}
    \includegraphics[width=\textwidth]{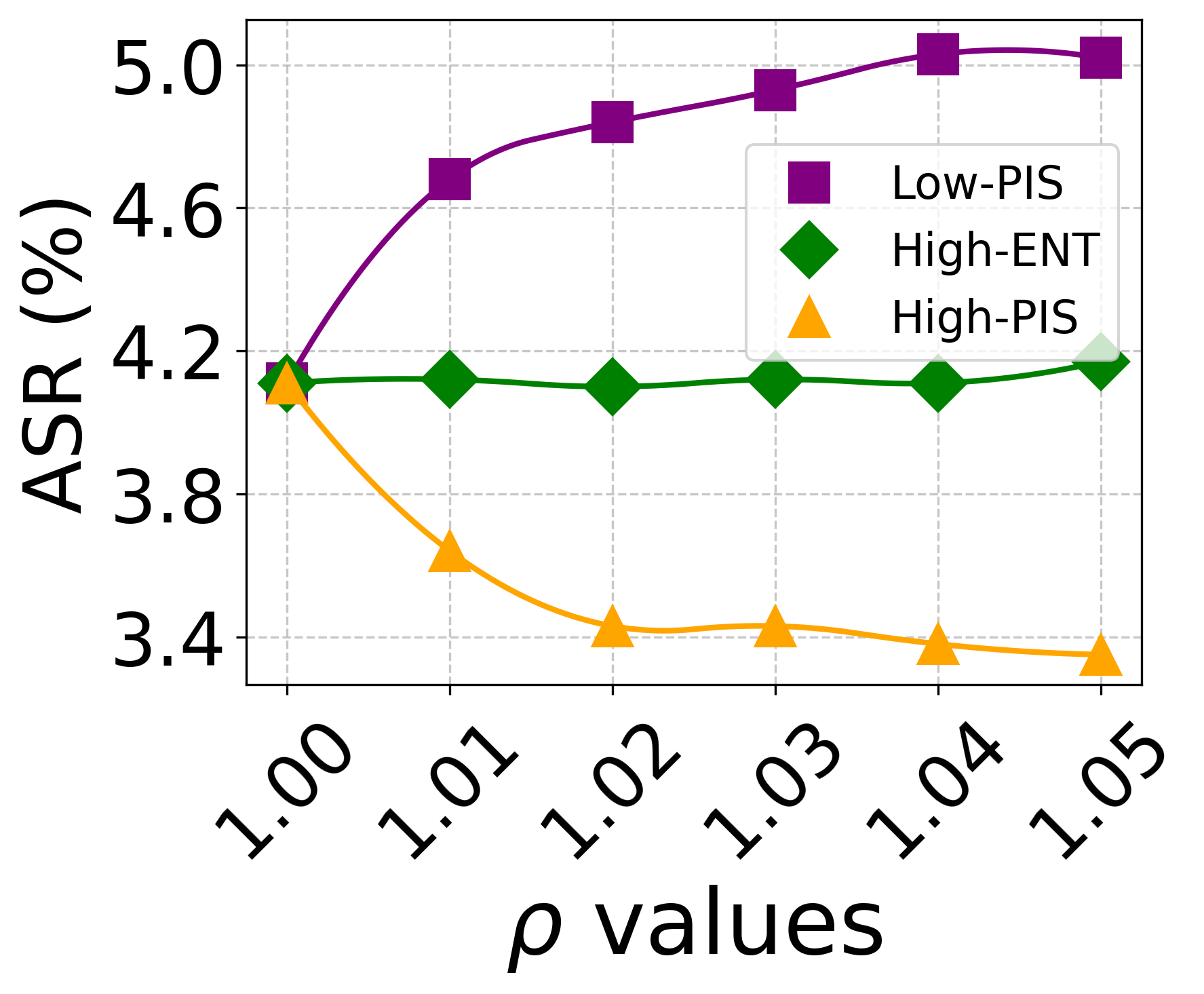}
  \end{minipage}
  \hfill
  \begin{minipage}[b]{0.24\textwidth}
    \includegraphics[width=\textwidth]{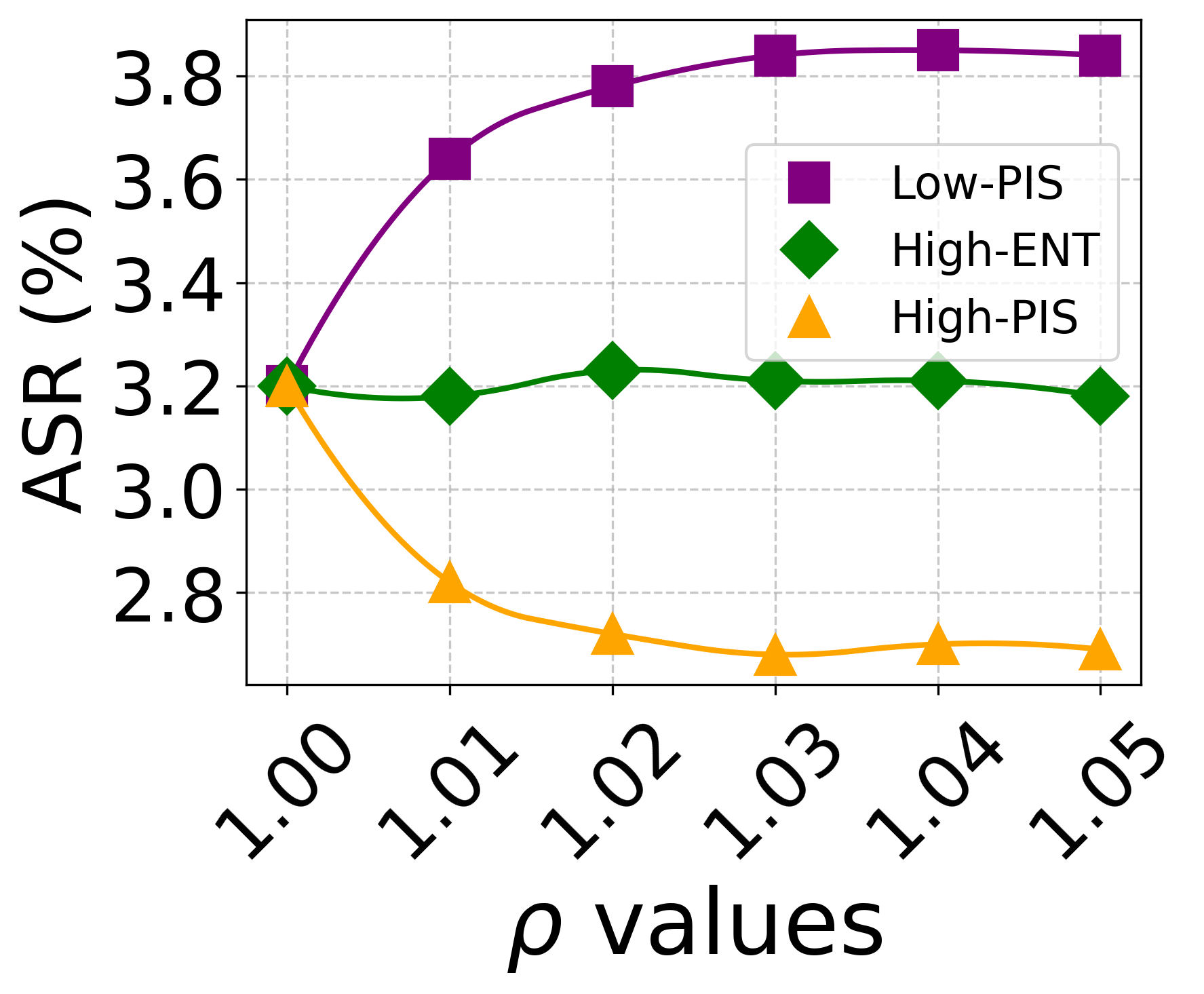}
  \end{minipage}
  \hfill
  \begin{minipage}[b]{0.24\textwidth}
    \includegraphics[width=\textwidth]{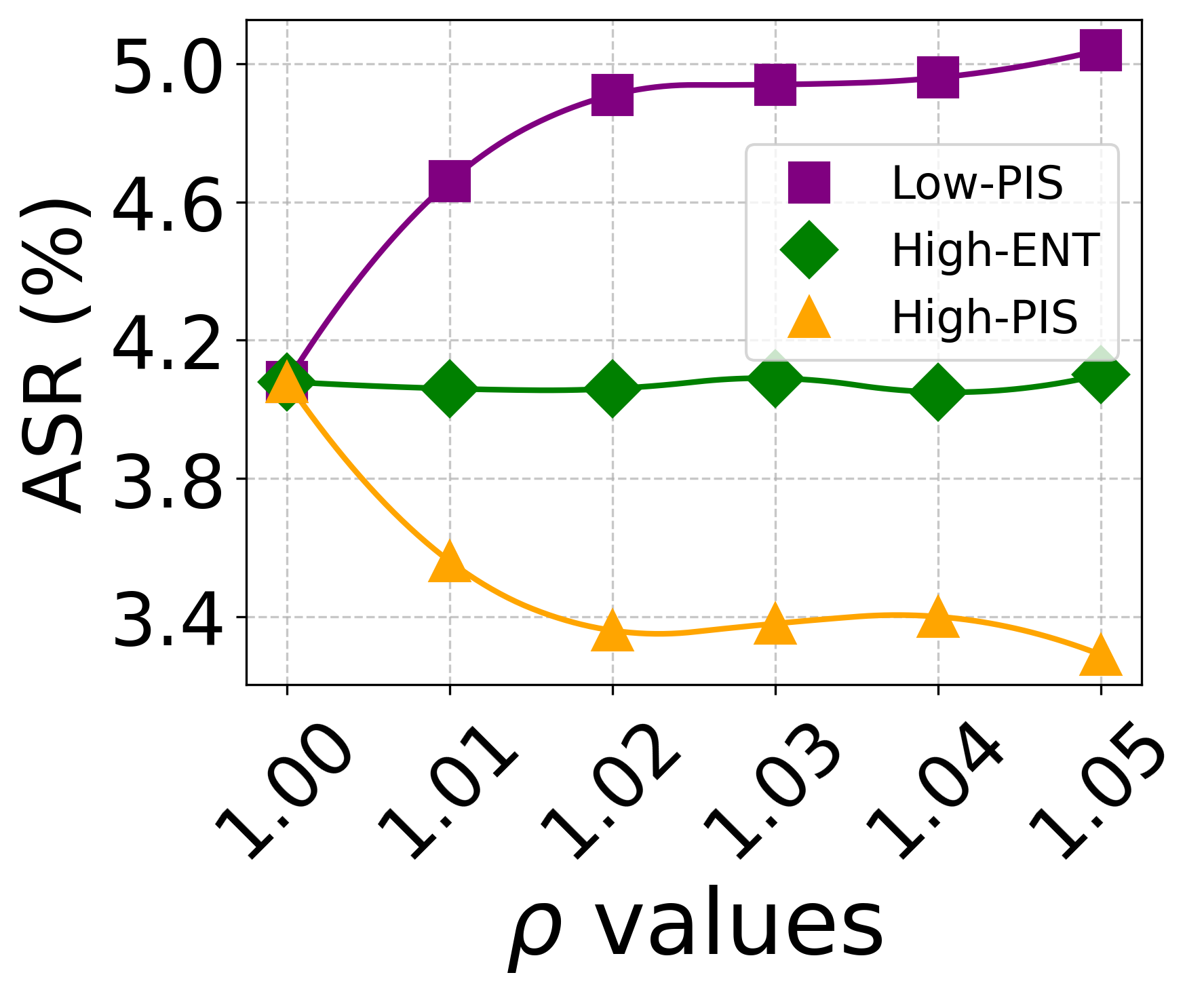}
  \end{minipage}
  \begin{minipage}[b]{0.24\textwidth}
    \includegraphics[width=\textwidth]{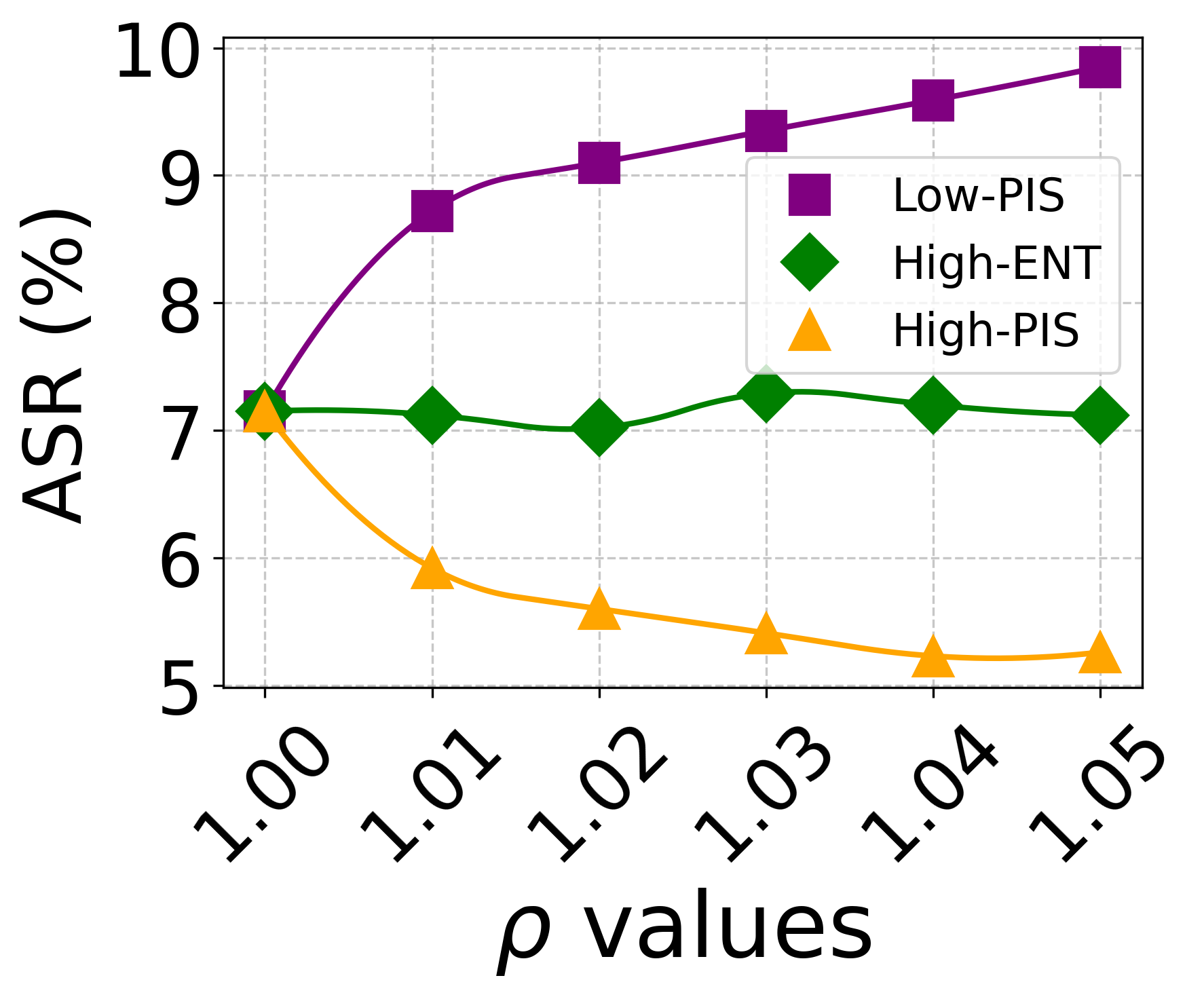}
  \end{minipage}
  \hfill
  \begin{minipage}[b]{0.24\textwidth}
    \includegraphics[width=\textwidth]{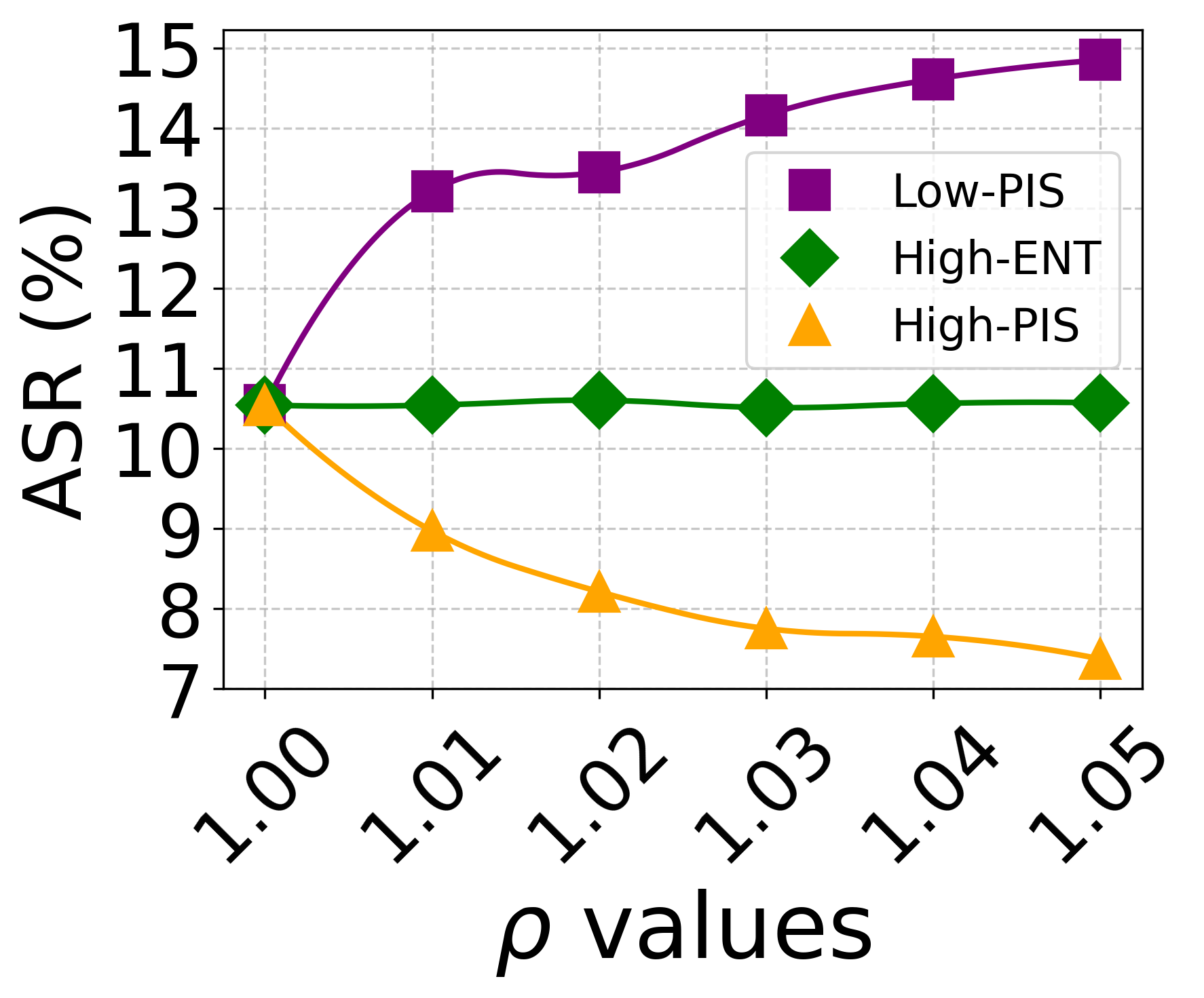}
  \end{minipage}
  \hfill
  \begin{minipage}[b]{0.24\textwidth}
    \includegraphics[width=\textwidth]{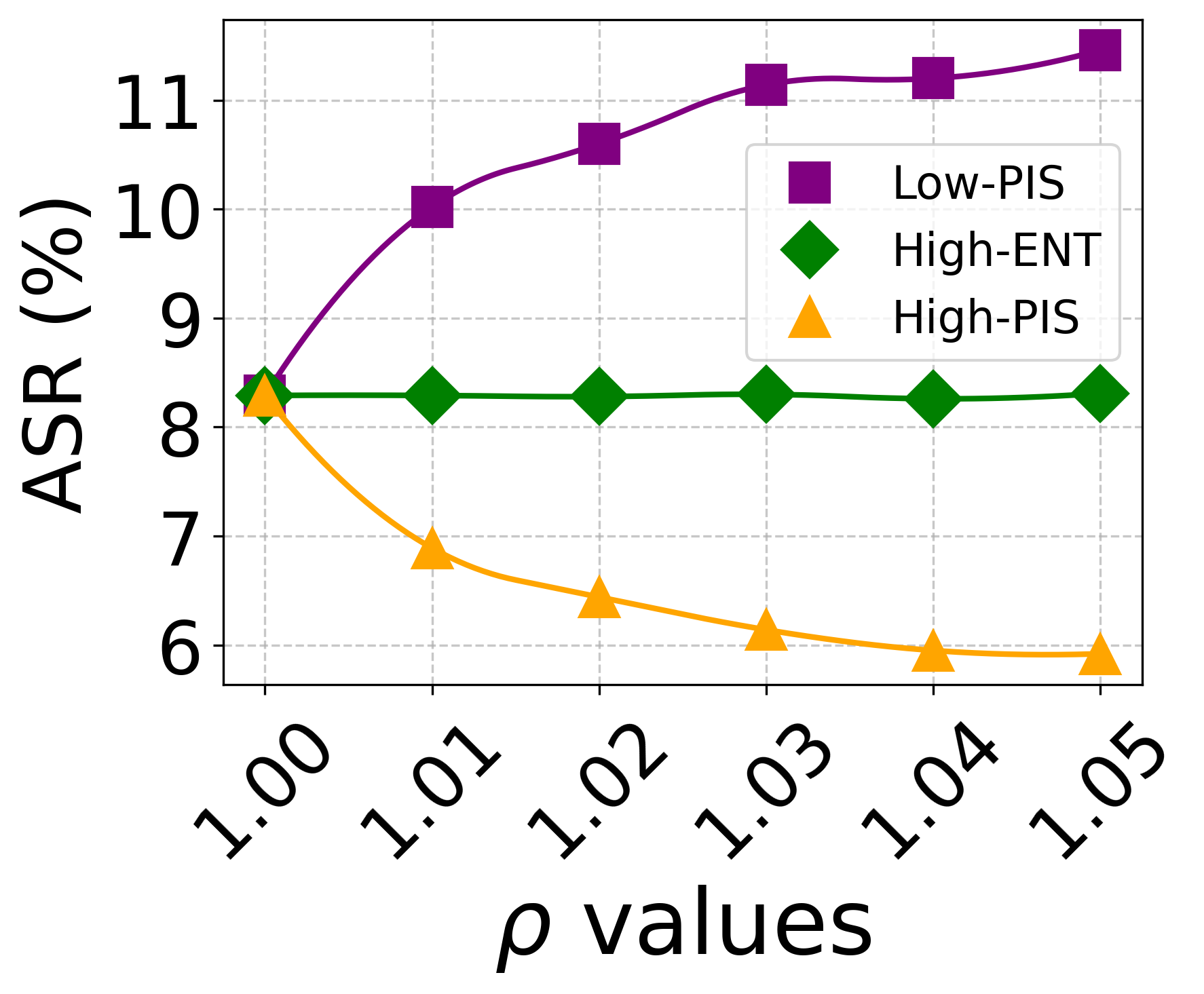}
  \end{minipage}
  \hfill
  \begin{minipage}[b]{0.24\textwidth}
    \includegraphics[width=\textwidth]{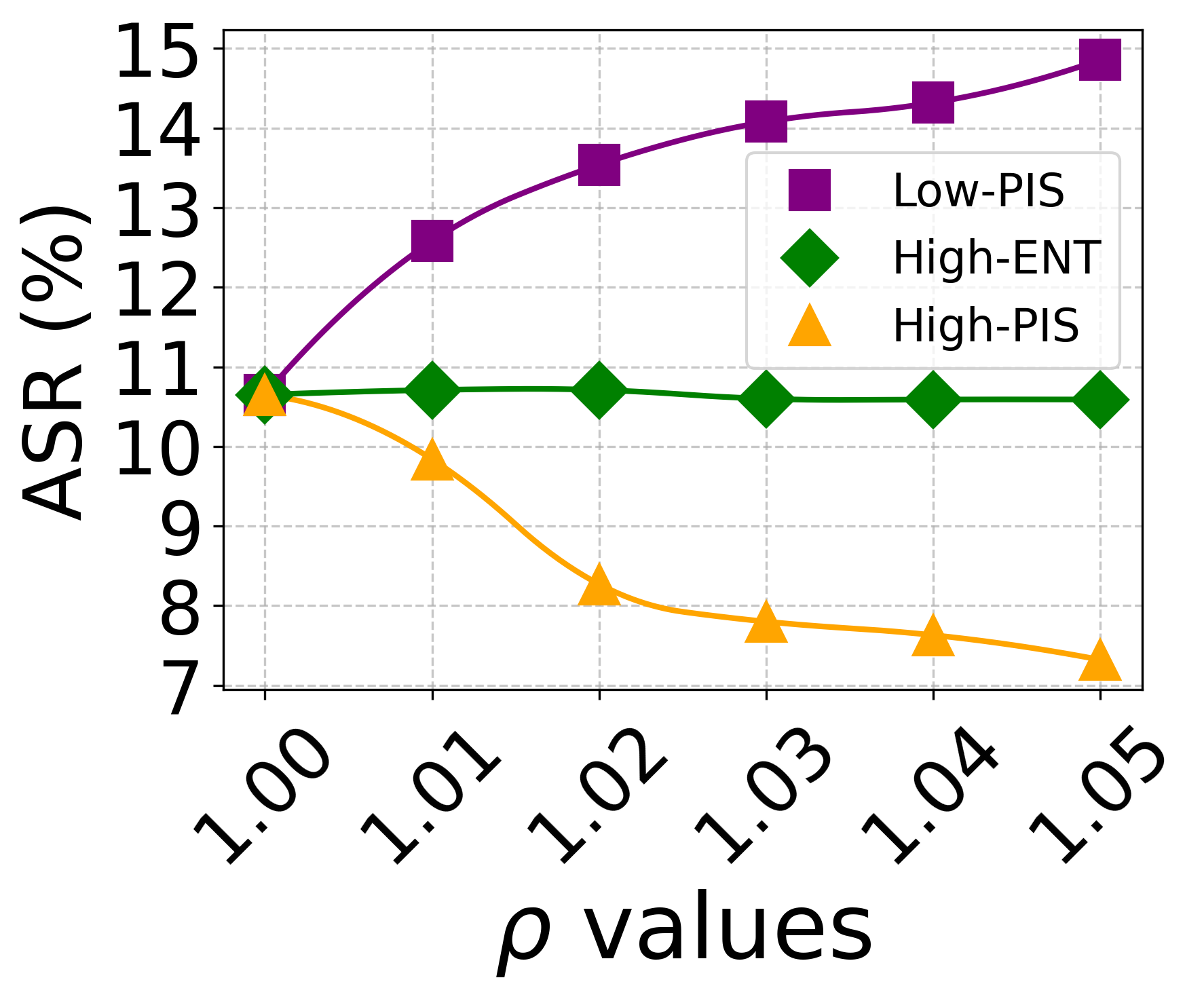}
  \end{minipage}
    \vspace{-6pt}
  \caption{BIA results on Adult, BMS-POS, Kosarak, and Gaussian datasets using the proposed F-OLH protocol instead of OLH. Top row is with $\varepsilon$ = 2, bottom row is with $\varepsilon$ = 3.}
  \label{fig:bia-folh}
  \vspace{-2pt}
\end{figure}

We again execute BIAs on the three subpopulations (High-ENT, Low-PIS, and High-PIS) and measure their ASRs. This time, as opposed to Section \ref{sec:impacts}, we assume that the whole population uses F-OLH instead of OLH. Hence, High-ENT, Low-PIS, and High-PIS subpopulations also consist of users who use F-OLH. Results with varying $\rho$ are shown in Figure \ref{fig:bia-folh}. 

A key take-away message from Figure \ref{fig:bia-folh} is that F-OLH is indeed effective in reducing the ASR differences between different subpopulations using the $\rho$ threshold. For example, ASRs of Low-PIS, High-ENT, and High-PIS users are all equal when $\rho$ = 1, indicating a fair outcome for all subpopulations. This is an intuitive result because $\rho$ = 1 enforces all users' hash functions to behave like $\mathcal{E}_\text{opt}$; therefore, there are no disparate impacts between different subpopulations. On the other hand, as $\rho$ is increased, the threshold becomes more relaxed. Consequently, we observe from Figure \ref{fig:bia-folh} that the ASR differences between the three subpopulations increase as $\rho$ is increased. This is also an intuitive result, since higher $\rho$ allows the use of more non-uniform hash functions, i.e., hash functions which diverge from $\mathcal{E}_\text{opt}$. As a result, unfairness in the user population increases, and the subpopulations start diverging from one another. While the differences between subpopulations are relatively small when $\rho$ = 1.01, they enlarge as $\rho$ is increased to 1.05. In fact, results with $\rho$ = 1.05 approach the results in Figure \ref{fig:bia}, implying that F-OLH converges to OLH, and it no longer provides much fairness benefit compared to OLH. Thus, we confirm that $\rho$ values closer to 1.0 better address unfairness with respect to BIA. 

\begin{figure*}[t]
    \centering
    \includegraphics[width=.9\textwidth]{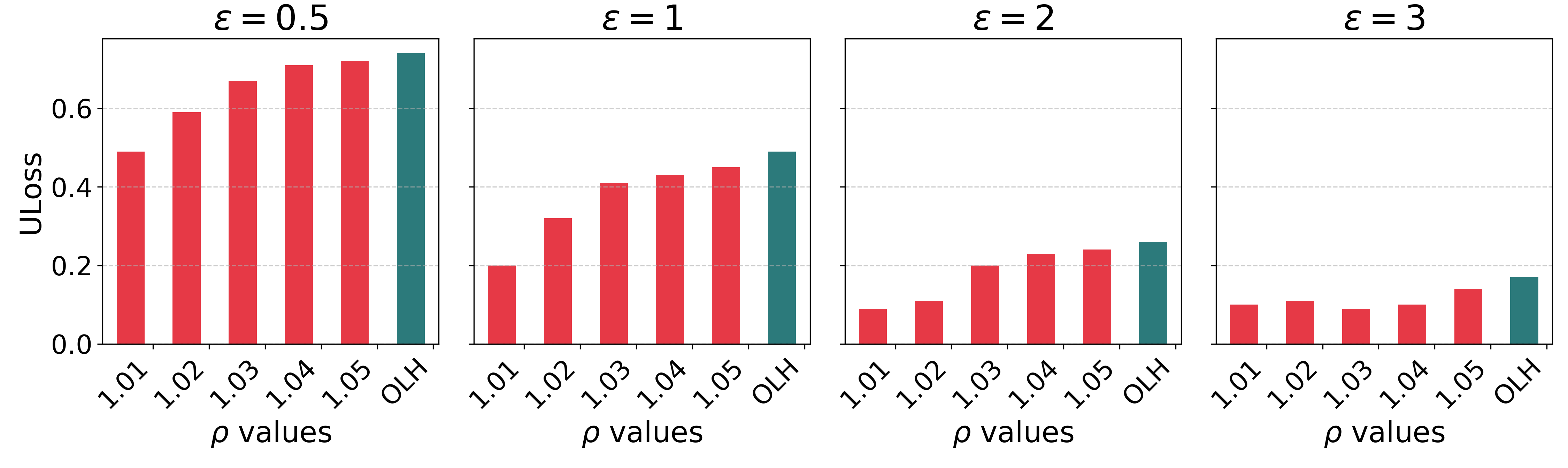}
    \includegraphics[width=.9\textwidth]{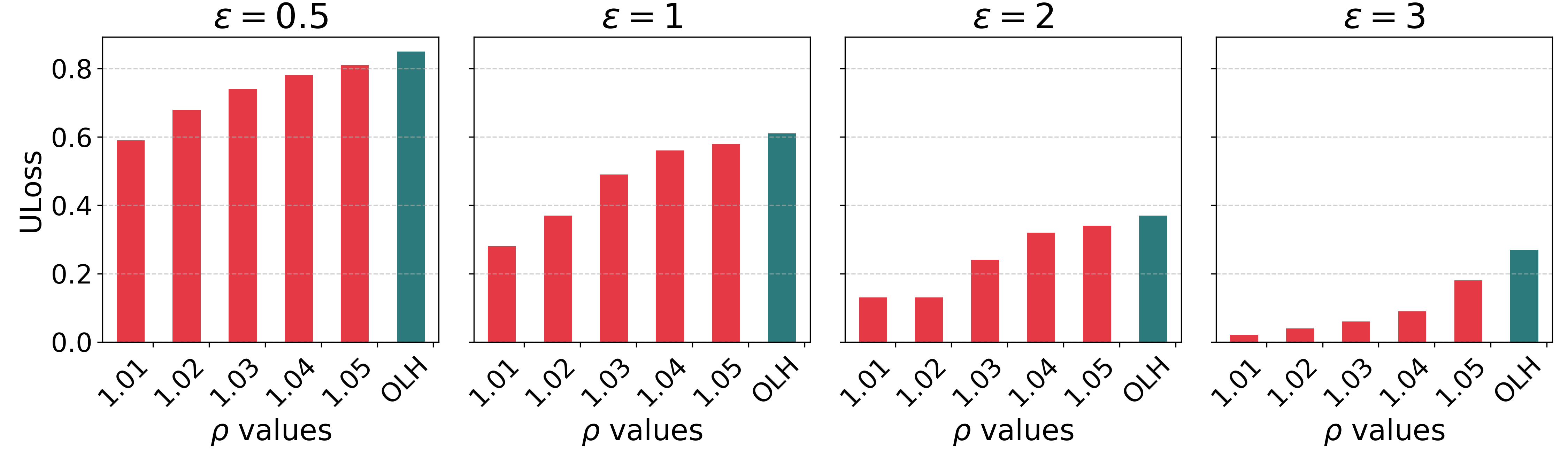}
    \vspace{-10pt}
    \caption{MGA results using the proposed F-OLH protocol. Top row is with the Kosarak dataset, bottom row is with the Gaussian dataset.}
    \vspace{-2pt}
    \label{fig:mga-folh}
\end{figure*}

In the next set of experiments, we compare the effectiveness of MGA on OLH and F-OLH under varying $\rho$ using the ULoss metric. Since previous results (reported in Figure \ref{fig:mga}) established that MGA is most effective on High-PIS by far, we focus our experiments here on the High-PIS subpopulation. The results are shown in Figure \ref{fig:mga-folh}. 


The results demonstrate that the effectiveness of MGA decreases significantly when F-OLH is used. Notably, when $\rho$ = 1.01, the effectiveness of MGA is nearly halved in F-OLH compared to OLH. As $\rho$ increases, F-OLH again gradually converges to OLH (similar to BIA results). The primary reason for this behavior is that under a strict $\rho$ constraint, the maximum preimage set size $P_u$ of a user ends up being much smaller in F-OLH compared to what it can be in OLH. As a result, MGA can manipulate fewer $\bar{v} \in T$, yielding smaller attack impact. It can also be observed from Figure \ref{fig:mga-folh} that under small $\varepsilon$ such as $\varepsilon$ = 0.5, ULoss is generally larger and it can remain large despite the use of F-OLH with strict $\rho$. The reason behind this trend is that the ULoss caused by LDP itself is larger when $\varepsilon$ is small. In other words, even if there was no attack like MGA, there would have been substantial ULoss when $\varepsilon$ is small, caused by LDP. Such fundamental (and natural) ULoss cannot be addressed by either OLH or F-OLH. 


Finally, we perform additional experiments on two real-world datasets (Kosa\-rak and BMS-POS) to analyze how the real frequency distribution $f(v)$ compares with frequencies after MGA attacks on OLH and F-OLH. We also perform additional experiments using Gaussian datasets with different $\sigma$. (The results of the latter experiment are similar to Figure \ref{fig:mga-folh}.) Due to the page limit, we provide the results and analysis of these experiments in the appendix. 

\subsection{Time Cost of F-OLH} 

In OLH, each user draws their hash function $H_u$ from $\mathcal{H}$ once. However, in F-OLH, the user continues drawing hash functions from $\mathcal{H}$ until a hash function satisfying the $\rho$ threshold is found. Thus, F-OLH has an additional time cost compared to OLH. In this section, we experimentally analyze this time cost. 

\begin{figure}[!t]
  \centering
  \begin{minipage}[b]{0.24\textwidth}
    \includegraphics[width=\textwidth]{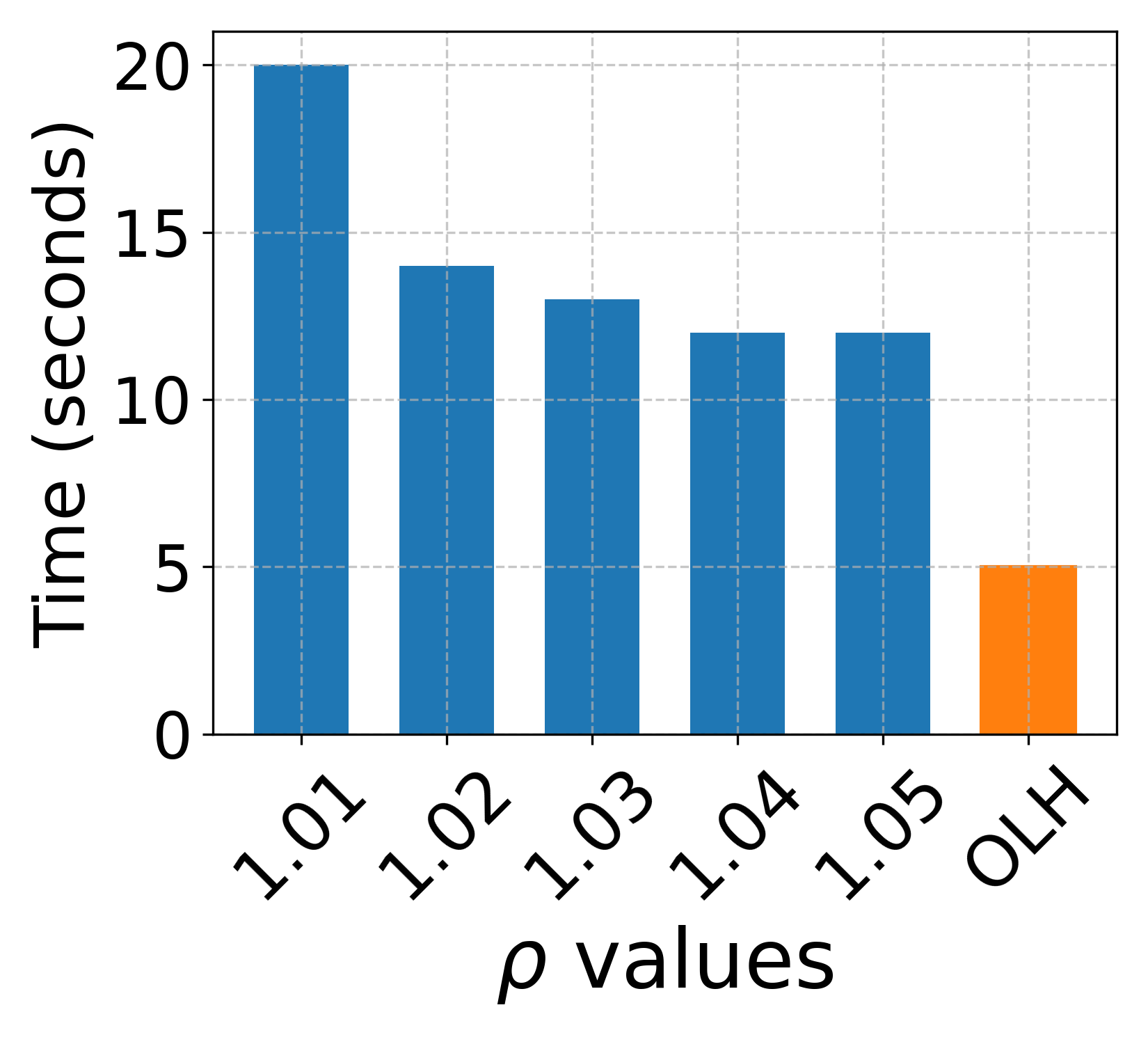}
  \end{minipage}
  \hfill
  \begin{minipage}[b]{0.24\textwidth}
    \includegraphics[width=\textwidth]{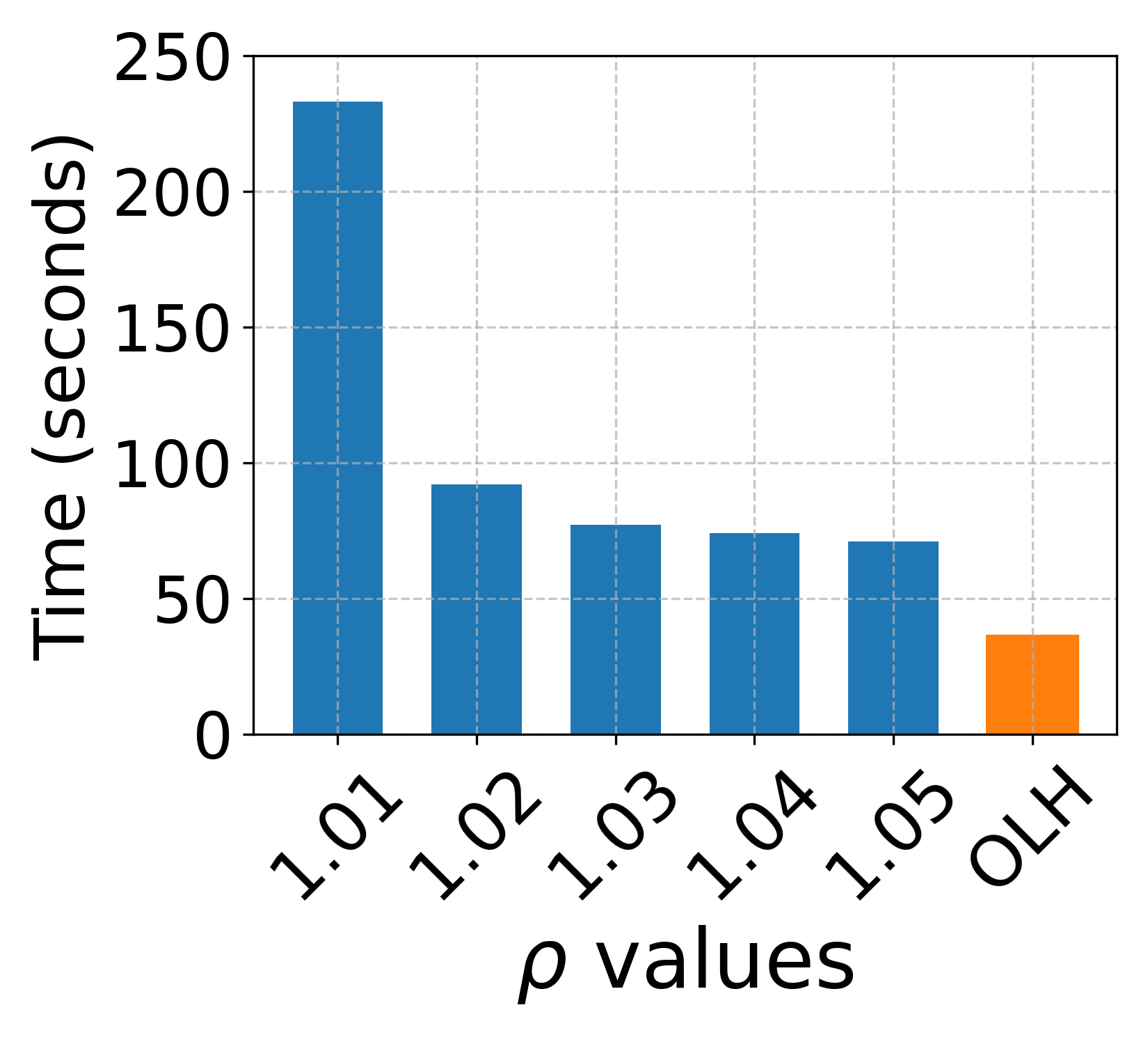}
  \end{minipage}
  \hfill
  \begin{minipage}[b]{0.24\textwidth}
    \includegraphics[width=\textwidth]{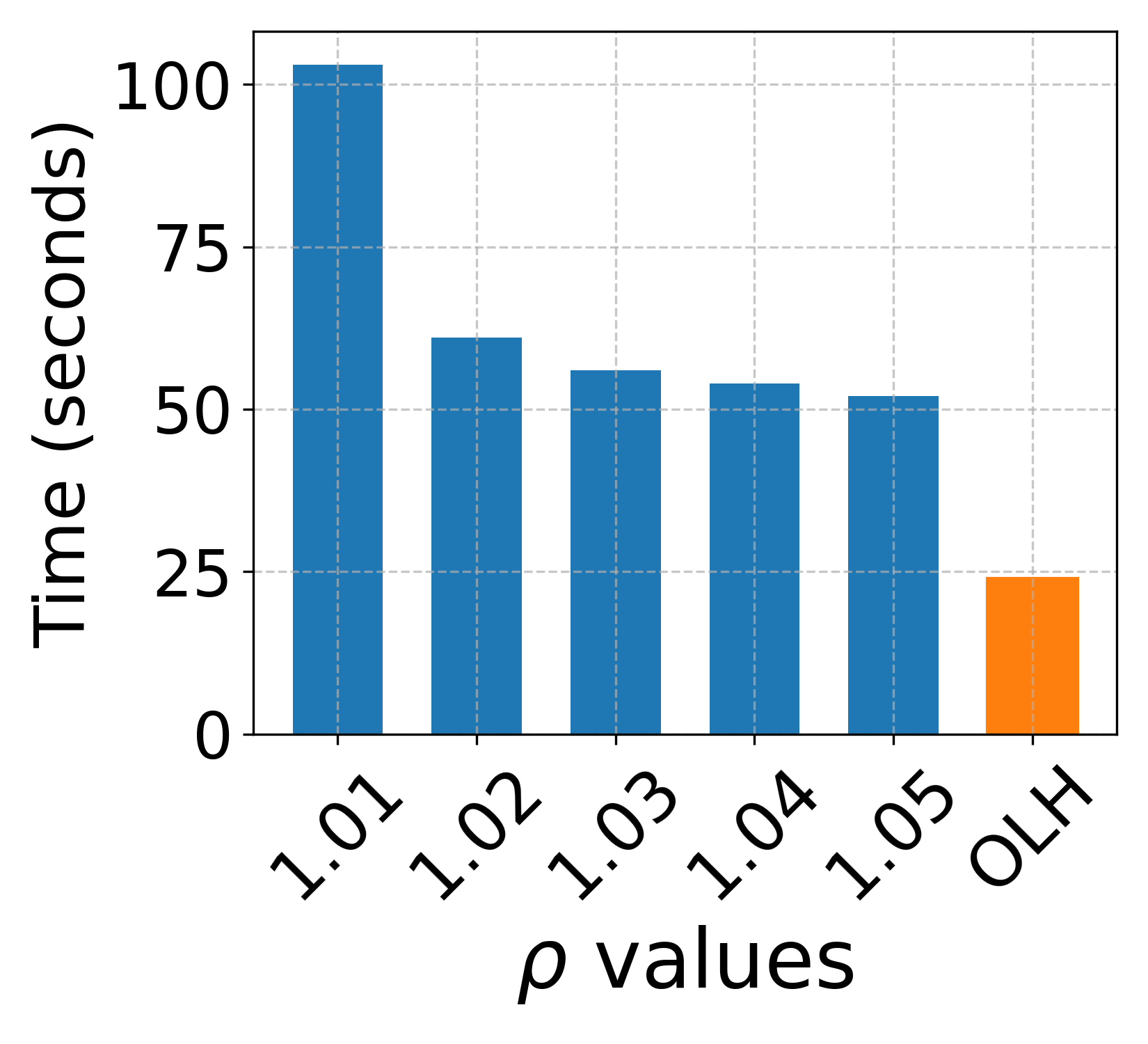}
  \end{minipage}
  \hfill
  \begin{minipage}[b]{0.24\textwidth}
    \includegraphics[width=\textwidth]{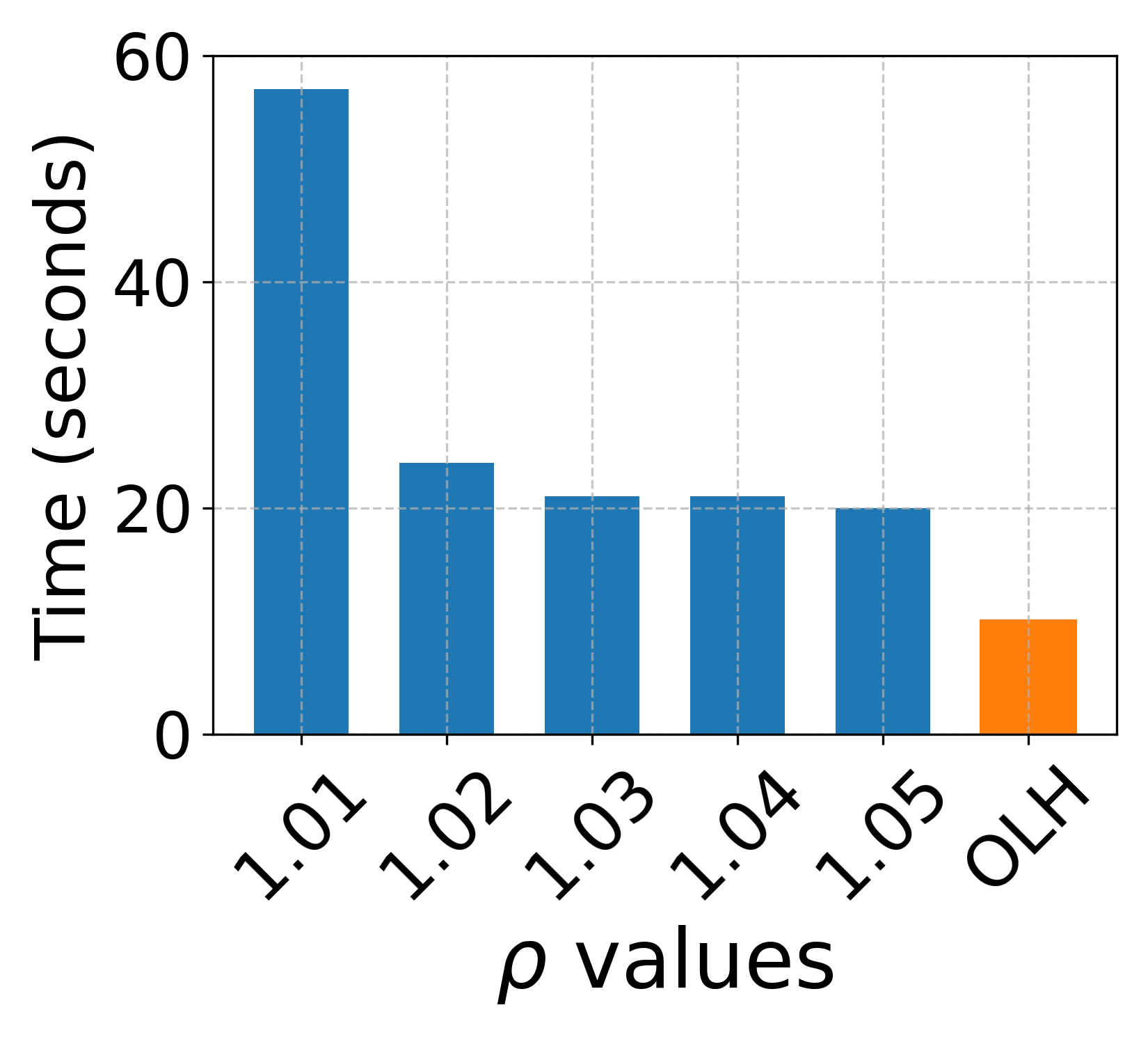}
  \end{minipage}
  \vspace{-10pt}
  \caption{Execution time comparison between F-OLH (blue bars) and OLH (orange bar) on Adult, BMS-POS, Kosarak, and Gaussian datasets.}
  \vspace{-4pt}
  \label{fig:time}
\end{figure}

In Figure \ref{fig:time}, we report the execution time comparison between OLH and F-OLH under varying $\rho$. In all plots, the value of $\varepsilon$ is fixed to $\varepsilon$ = 2. We observe that when $\rho$ is small, execution times are higher. This is because F-OLH needs to spend more time to find a hash function that satisfies the strict $\rho$ threshold, leading to increased computational overhead. As $\rho$ increases, it is easier to find a hash function that satisfies the threshold; therefore, execution times of F-OLH begin to converge to those of OLH. However, execution times of F-OLH always remain higher than OLH due to the additional computation in F-OLH: even when the $\rho$ constraint is relaxed, F-OLH still needs to call Algorithm \ref{alg:entropy} to compute $\mathcal{E}_{\text{comp}}$ and verify compliance with the $\rho$ threshold. This computation, which exists in F-OLH but does not exist in OLH, causes F-OLH's execution times to always be higher than OLH.

\begin{figure}[!t]
  \centering
  \hfill
  \begin{minipage}[b]{0.45\textwidth}
    \includegraphics[width=\textwidth]{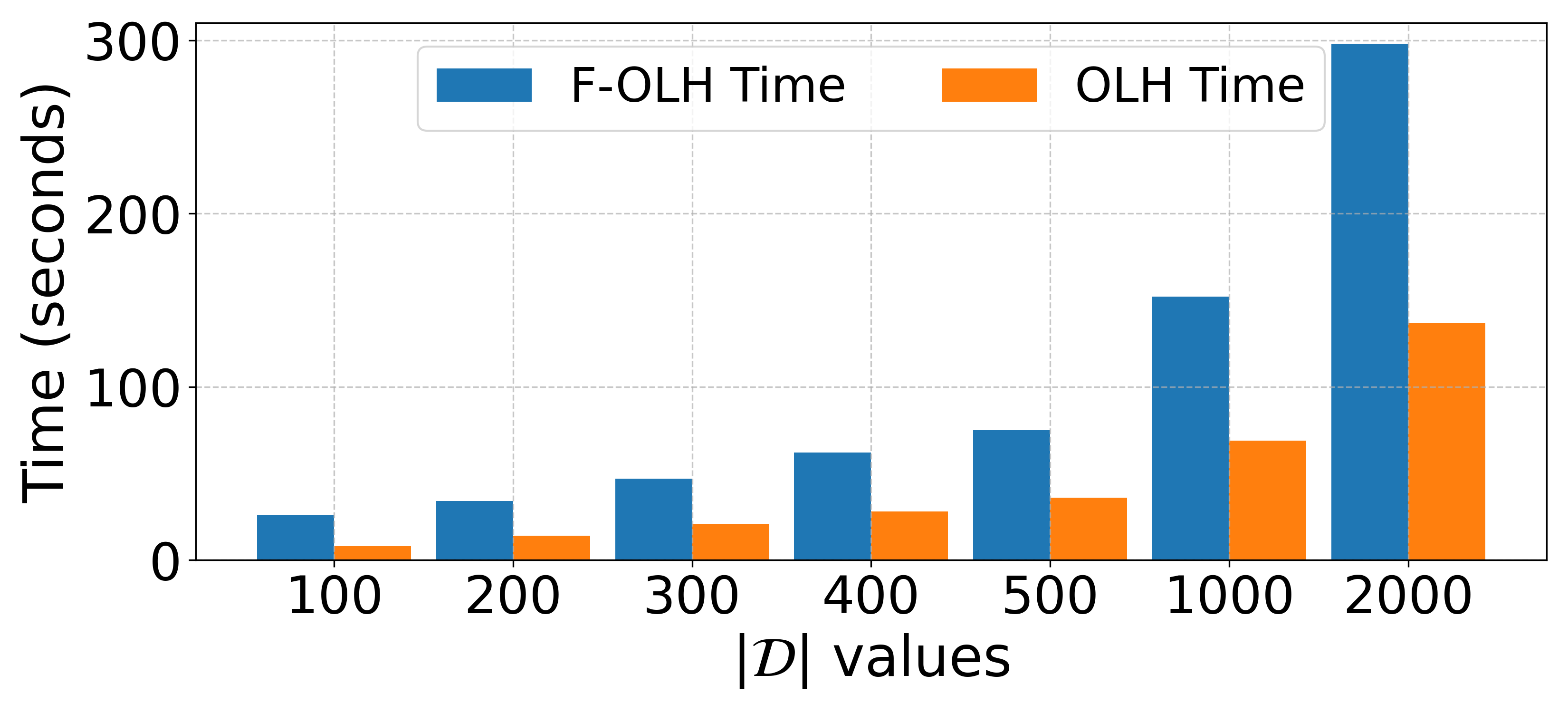}
  \end{minipage}
  \hfill
  \begin{minipage}[b]{0.45\textwidth}
    \includegraphics[width=\textwidth]{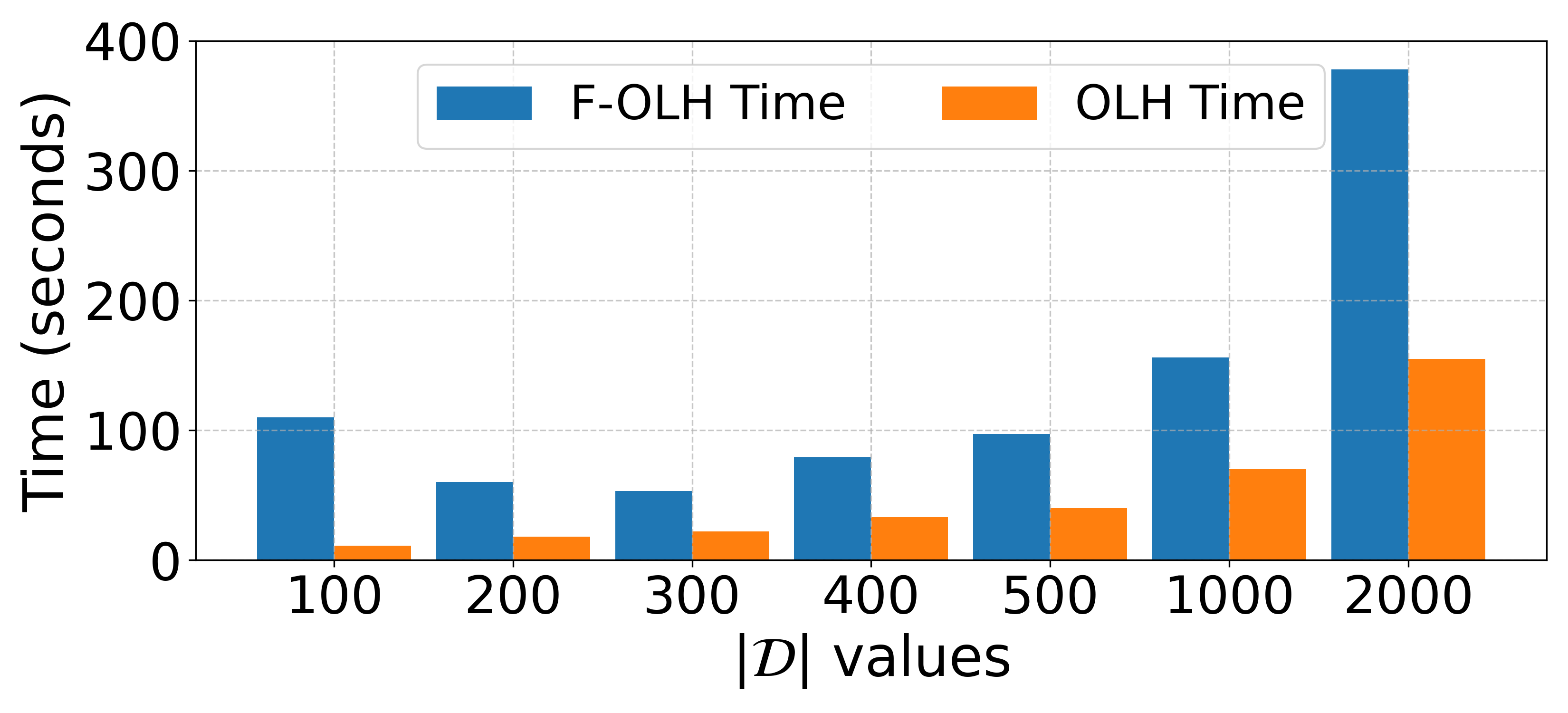}
  \end{minipage}
  \hfill
  \vspace{-10pt}
  \caption{Execution time comparison between F-OLH and OLH on Uniform datasets with varying domain sizes $|\mathcal{D}|$. $g$ = 5 on the left, $g$ = 10 on the right.}
  \label{fig:uni-time}
  \vspace{-2pt}
\end{figure}

Next, we use the Uniform datasets to measure how the domain size $|\mathcal{D}|$ affects the execution times of OLH and F-OLH. The results of this experiment are shown in Figure \ref{fig:uni-time}. We fixed $g$ = 5 for the plot on the left (corresponding to approximately $\varepsilon$ = 1.5) and $g$ = 10 for the plot on the right (corresponding to approximately $\varepsilon$ = 2.2). In both figures, $\rho$ = 1.01. Again, we observe that in general, the execution times of F-OLH are higher than OLH. As the domain size increases, the execution time for both protocols increases. A notable outlier, however, is the case when $g$ = 10 and $|\mathcal{D}|$ = 100. In this case, the execution time is higher than many other $|\mathcal{D}|$. We found the reason behind this outlier is that it is challenging for F-OLH to find a suitable hash function which ensures a near-uniform distribution (so that $\rho$ = 1.01 is satisfied) for a relatively small $|\mathcal{D}|$ and large $g$ = 10. Many iterations of the while loop in Algorithm \ref{alg:FOLH} are necessary to find a suitable hash function, thereby increasing execution time.  

Based on these results, we arrive at the following conclusions: While larger $|\mathcal{D}|$ usually yields higher execution times, the values of $\rho$ and $g$ (which is tied to $\varepsilon$) are also important, especially in F-OLH. This is because $\rho$, $g$, and $|\mathcal{D}|$ all affect how difficult it is for the users to find a hash function that satisfies the $\rho$ constraint. We also note that the execution times reported in Figures \ref{fig:time} and \ref{fig:uni-time} are the results of simulating all users on one CPU, sequentially (one user after the other). In the real world, all users will be executing OLH or F-OLH in parallel on their own devices. Thus, the real-world execution time costs of the protocols will be close to the execution times reported in Figures \ref{fig:time} and \ref{fig:uni-time} divided by $|\mathcal{U}|$. Hence, the execution time for each individual user will remain in the order of milliseconds, both in OLH and F-OLH. This shows that in a practical implementation, replacing OLH by F-OLH will not lead to a noticeable execution time problem for the users.

In the previous section, we established that selecting $\rho$ close to 1 improves fairness and resilience to both BIA and MGA attacks. In this section, we established that it also has the adverse effect of increasing execution time. In the appendix, we provide guidance regarding how to select $\rho$ in a practical setting.

\vspace{-2pt}
\section{Related Work} \label{sec:RelatedWork}

Since our paper aims to expose and mitigate hash-induced unfairness in LDP protocols in terms of privacy and poisoning attacks, we survey related literature in three categories: (i) privacy attacks in LDP, (ii) poisoning attacks in LDP, (iii) intersection of fairness and LDP.

\textbf{Privacy attacks in LDP} typically aim to infer sensitive user information or assess practical privacy guarantees of LDP protocols against specific adversaries. Murakami et al.~\cite{murakami2020toward} examined re-identification risks in LDP. Gadotti et al.~\cite{gadotti2022pool} introduced pool inference attacks to determine if a user's true value belongs to a certain pool based on the LDP protocol output. Gursoy et al.~\cite{gursoy2022adversarial} analyzed LDP protocols using a Bayesian adversary model, and later extended this model to longitudinal data collection \cite{gursoy2024longitudinal}. Arcolezi et al.~\cite{arcolezi2023risks} studied inference risks stemming from multidimensional data collection. Finally, Arcolezi and Gambs \cite{arcolezi2024revealing} proposed LDP-Auditor for empirically estimating the privacy loss of LDP mechanisms. Following the literature, this paper uses the Bayesian adversary model, which was proposed in \cite{gursoy2022adversarial} and later extended and used in \cite{arcolezi2025revisiting,arcolezi2023risks,gursoy2024longitudinal}.

\textbf{Poisoning attacks in LDP.} In poisoning attacks, malicious users poison the LDP protocol to manipulate the statistics recovered by the data collector. The vulnerability of LDP protocols to poisoning was shown by Cheu et al.~\cite{cheu2021manipulation} and Cao et al.~\cite{mga}. Poisoning attacks targeting mean and variance estimation were proposed in \cite{li2023fine}. Li et al.~\cite{li2024robustness} studied the robustness of LDP protocols designed for numerical attributes. Wu et al.~\cite{wu2022poisoning} proposed poisoning attacks for key-value data, Imola et al.~\cite{imola2022robustness} studied poisoning attacks on graph analysis, and Zheng et al.~\cite{zheng2024data} studied poisoning attacks against LDP-based privacy-preserving crowdsensing. Since we are working directly with LDP protocols (i.e., not an application such as key-value data or crowdsensing), in this paper, we use MGA from the literature \cite{mga} as the fundamental poisoning attack. 

\textbf{Fairness and LDP.} There also exist some recent works at the intersection of LDP and fairness. Chen et al.~\cite{chen2022fairness} propose FairSP to achieve fair machine learning predictions while some sensitive attributes are protected with LDP. Mozannar et al.~\cite{mozannar2020fair} study fair learning under privatized demographic data. Arcolezi et al.~\cite{arcolezi2023local} and Makhlouf et al.~\cite{makhlouf2024impact} study the impacts of perturbing multidimensional records with LDP on fairness. However, all of these works consider fairness in terms of classification models (i.e., supervised machine learning) built from privatized data. None of the previous works consider fairness from the perspective of increased (or decreased) vulnerability to privacy and poisoning attacks. To the best of our knowledge, we are the first to study the unfairness of LDP protocols in terms of privacy protection and vulnerability to poisoning. 

\section{Conclusion}

In this paper, we identified and analyzed hash-induced unfairness in LDP protocols. We demonstrated that users employing non-uniform hash functions experience significantly different vulnerabilities to BIA and MGA attacks, despite using the same protocol and privacy parameters. To mitigate this disparity, we proposed Fair-OLH (F-OLH), a variant of OLH that enforces an entropy-based fairness constraint on hash function selection. Our experiments showed that F-OLH substantially reduces disparities among users, improving fairness in both Bayesian privacy protection (BIA) and poisoning attacks (MGA), while maintaining acceptable time cost. These findings highlight the critical need to account for hash function behavior in LDP protocols. In future work, we plan to explore broader notions of fairness, explore more efficient methods for fair hash function selection, and propose fair variants of protocols other than OLH. 

\vspace{-4pt}
\subsubsection{Acknowledgements} 

This study was supported by The Scientific and Technological Research Council of Turkiye (TUBITAK) under grant number 123E179. The authors thank TUBITAK for their support.

\clearpage
\newpage

\bibliographystyle{splncs04}
\bibliography{references}

\appendix

\section{Additional Experiments with MGA}

\begin{figure}[!t]
  \centering
  \begin{minipage}[b]{0.24\textwidth}
    \includegraphics[width=\textwidth]{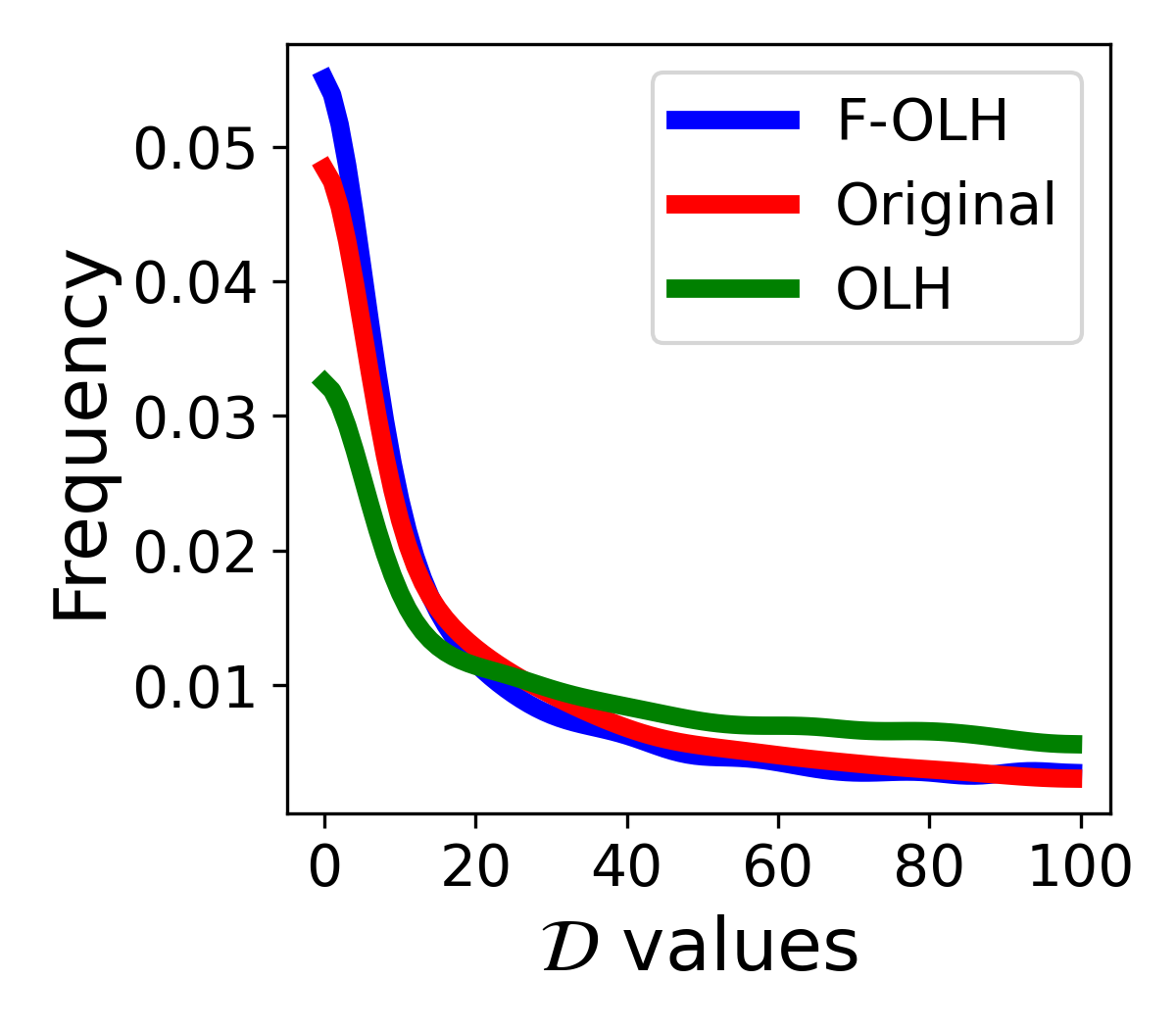}
  \end{minipage}
  \hfill
  \begin{minipage}[b]{0.24\textwidth}
    \includegraphics[width=\textwidth]{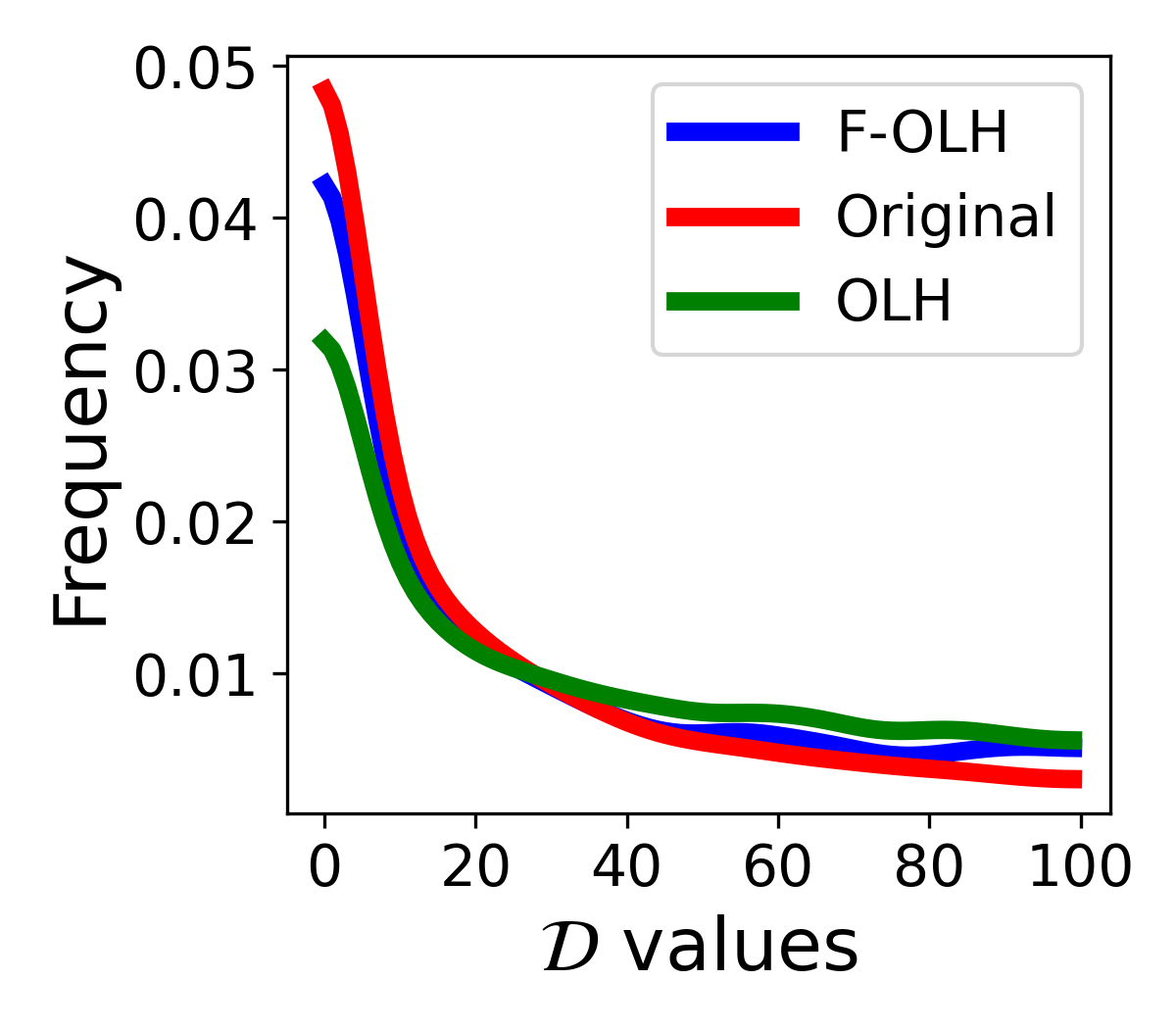}
  \end{minipage}
  \hfill
  \begin{minipage}[b]{0.24\textwidth}
    \includegraphics[width=\textwidth]{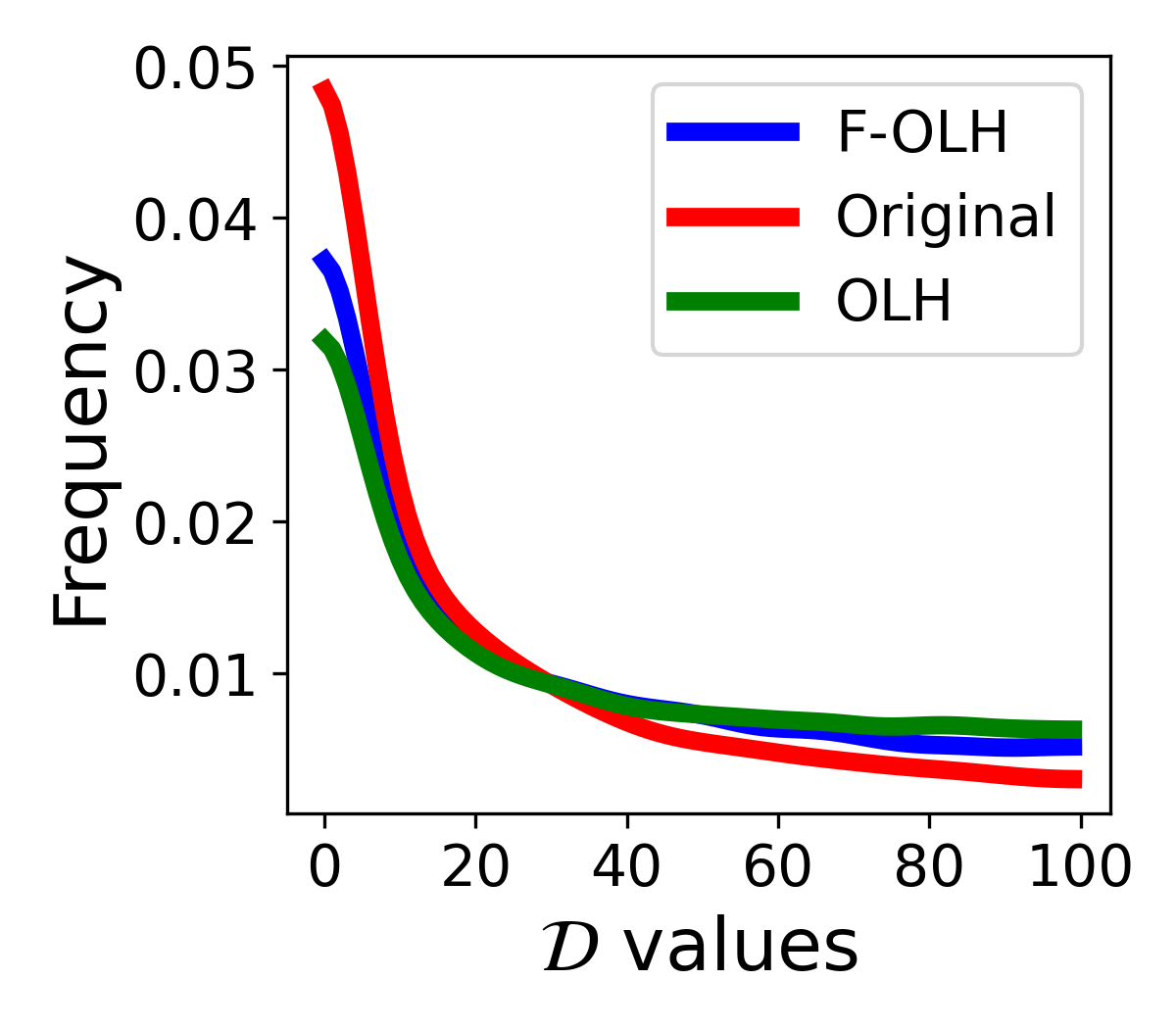}
  \end{minipage}
  \hfill
  \begin{minipage}[b]{0.24\textwidth}
    \includegraphics[width=\textwidth]{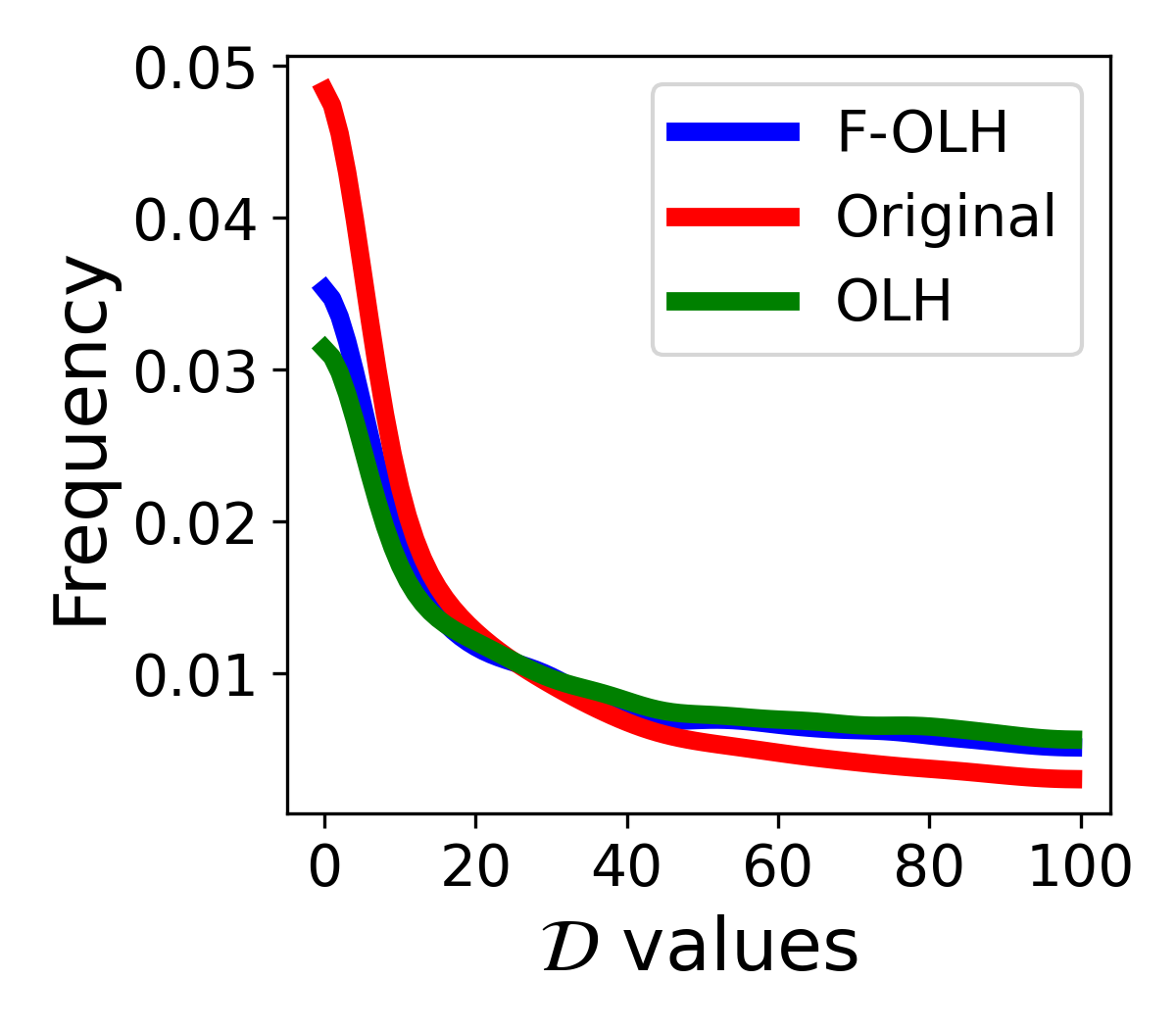}
  \end{minipage}
  \begin{minipage}[b]{0.24\textwidth}
    \includegraphics[width=\textwidth]{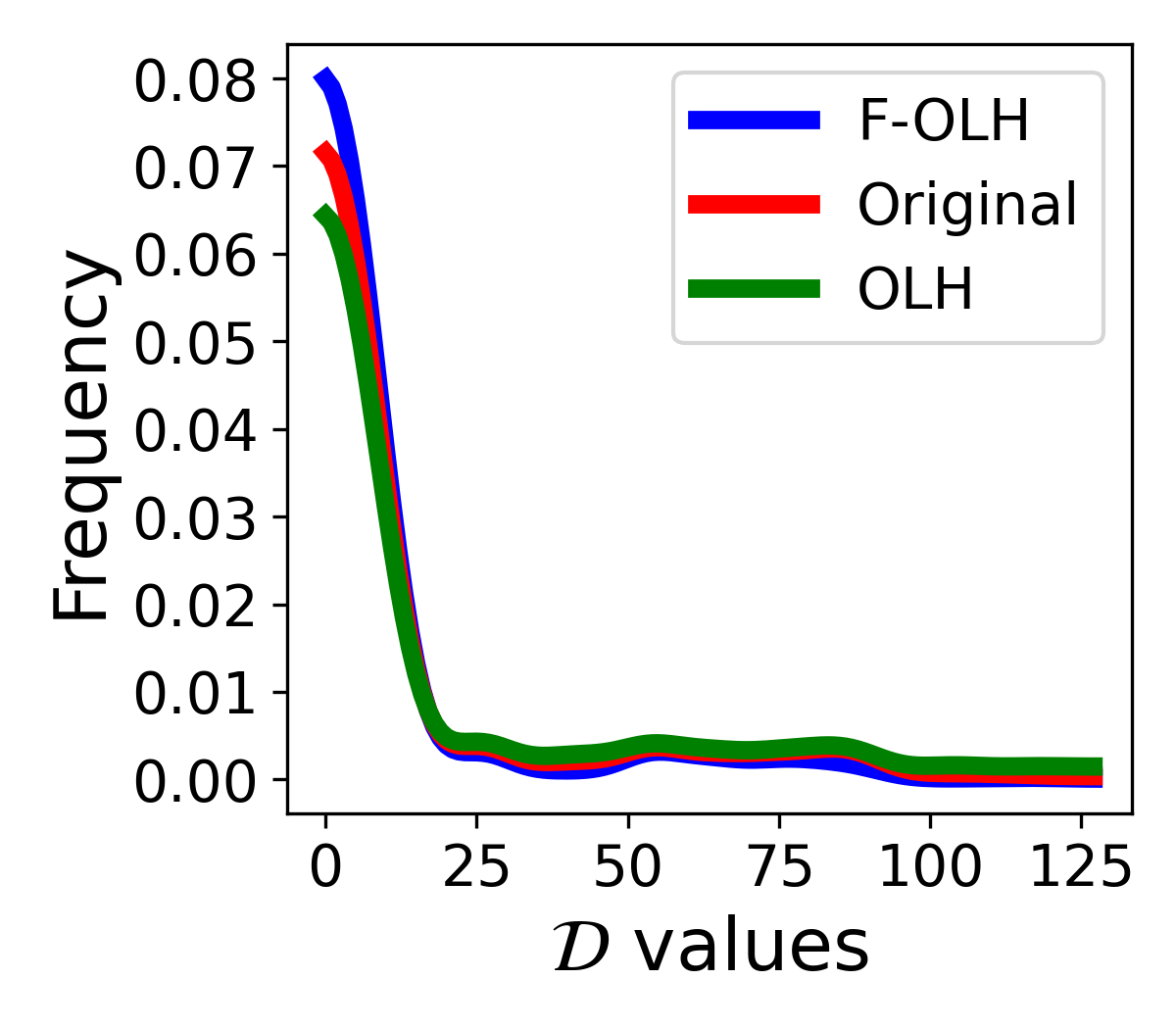}
  \end{minipage}
  \hfill
  \begin{minipage}[b]{0.24\textwidth}
    \includegraphics[width=\textwidth]{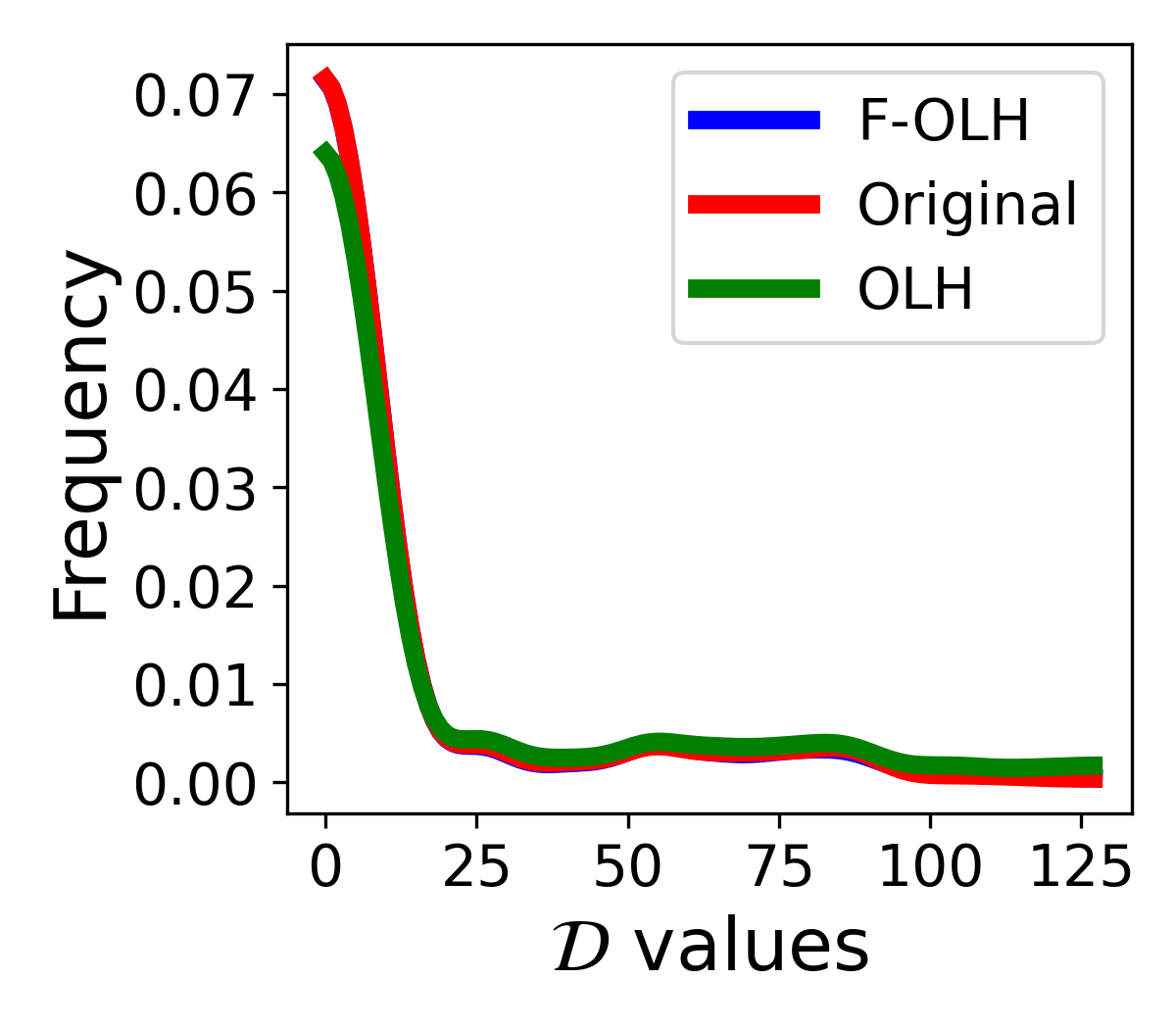}
  \end{minipage}
  \hfill
  \begin{minipage}[b]{0.24\textwidth}
    \includegraphics[width=\textwidth]{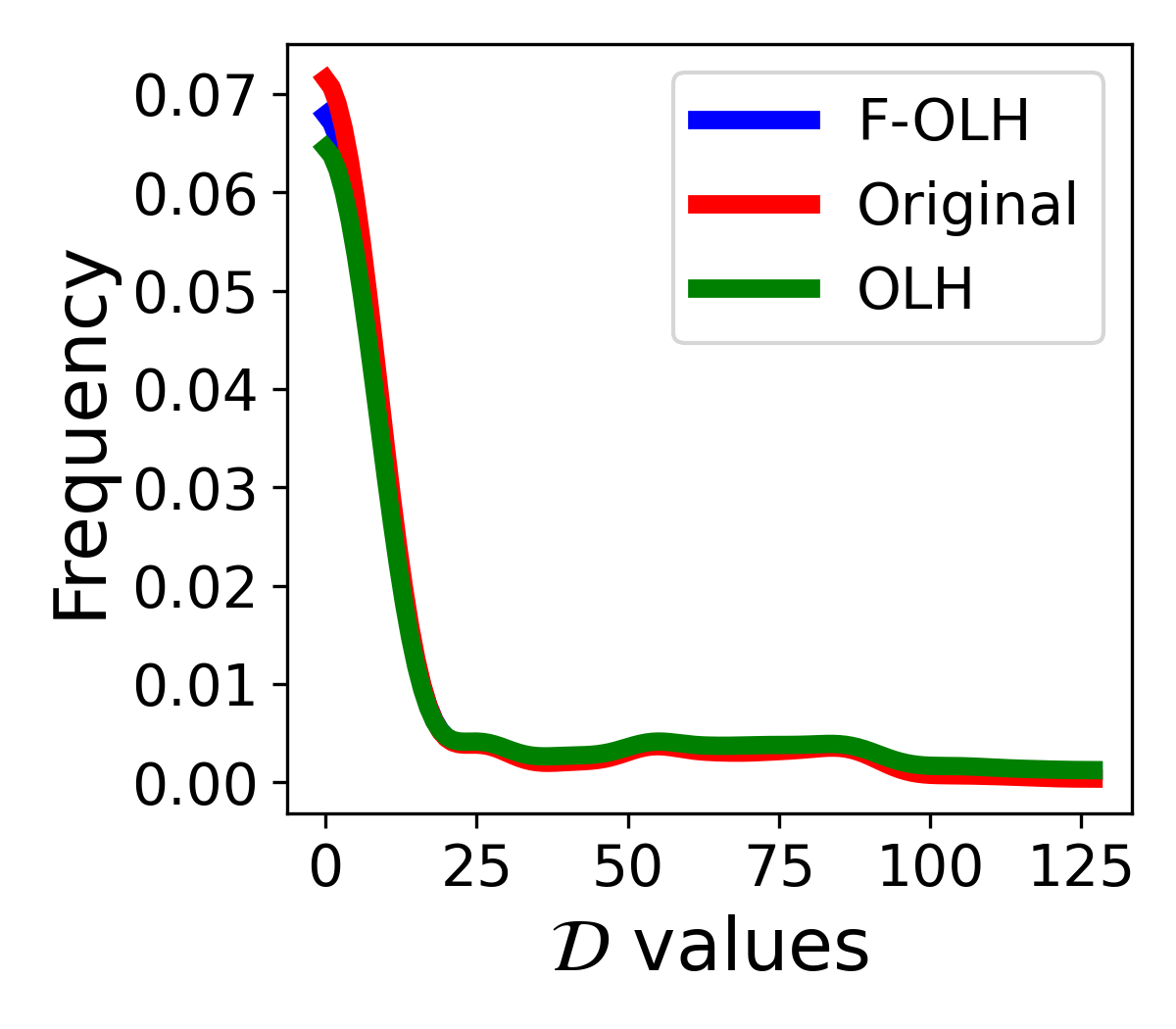}
  \end{minipage}
  \hfill
  \begin{minipage}[b]{0.24\textwidth}
    \includegraphics[width=\textwidth]{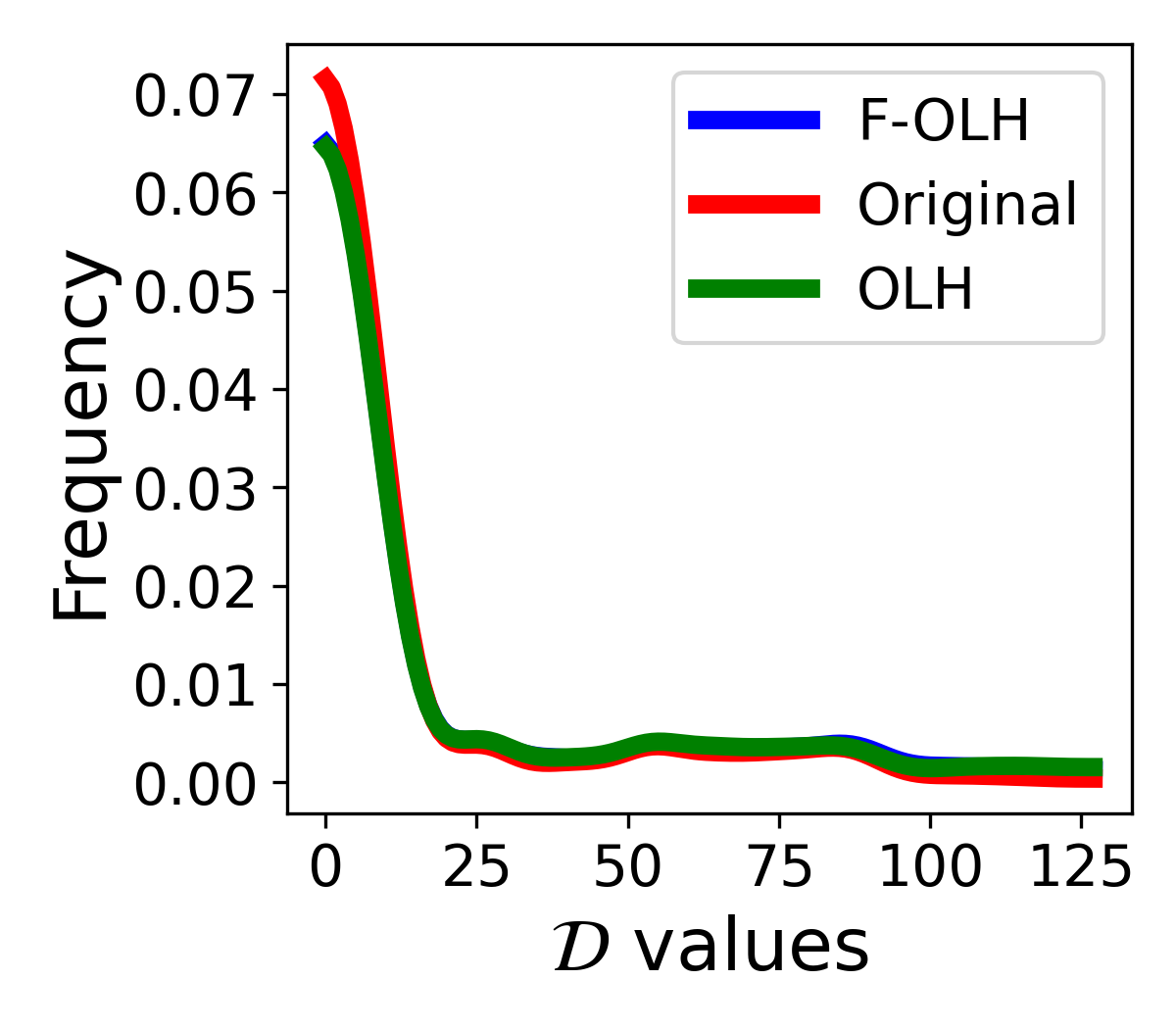}
  \end{minipage}
    \vspace{-6pt}
  \caption{Real frequencies $f(v)$ of items $v \in \mathcal{D}$ (red curve) versus estimated frequencies $\tilde{f}(v)$ under MGA attack with OLH (green curve) and F-OLH (blue curve). $\varepsilon$ = 0.5 is used in all plots, $\rho$ values are $\rho$ = 1.01, 1.02, 1.03, 1.04 from left to right. Top row: BMS-POS dataset, bottom row: Kosarak dataset.}
  \label{fig:mga-freqs}
\end{figure}

In Figure \ref{fig:mga-freqs}, we present the results of the MGA attack on BMS-POS and Kosarak, using 2000 malicious users for BMS-POS and 1000 malicious users for Kosarak. The increase in malicious users for BMS-POS is because BMS-POS consists of 400000 users, almost twice the size of Kosarak. The three lines in the plots represent the original frequencies without LDP, the estimated frequencies after MGA on F-OLH, and the estimated frequencies after MGA on OLH.

We observe from the plots that OLH exhibits an interesting behavior: frequencies of frequent values (high $f(v)$) are underestimated, while frequencies of infrequent values (low $f(v)$) are overestimated. However, F-OLH performs better than OLH, i.e., its overestimation and underestimation amounts are lower than those of OLH. This is because of the uniformity of hash functions used in F-OLH. Especially when $\rho$ is small, F-OLH effectively ``scales up'' the frequencies, causing frequent values to be estimated higher whereas infrequent values are estimated lower. This characteristic fixes the underestimations and overestimations of OLH, thereby acting as a natural defense mechanism against MGA. It also allows F-OLH to maintain a frequency distribution which more closely resembles the original distribution, thereby yielding lower ULoss than OLH.

\begin{figure}[!t]
\centering
    \includegraphics[width=.85\textwidth]{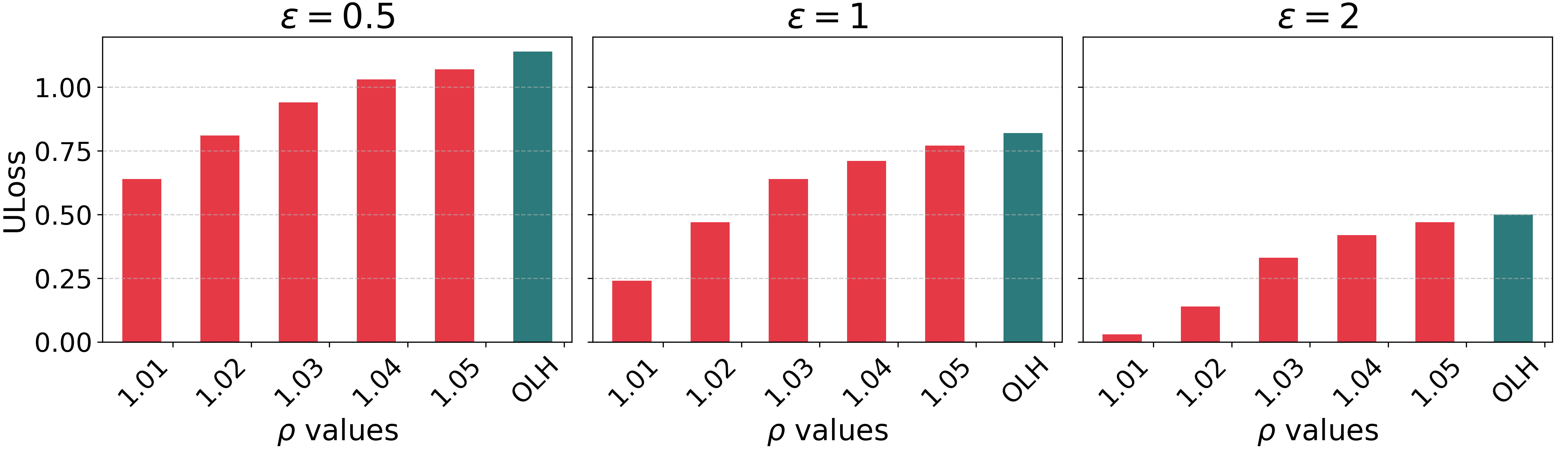}
    \includegraphics[width=.85\textwidth]{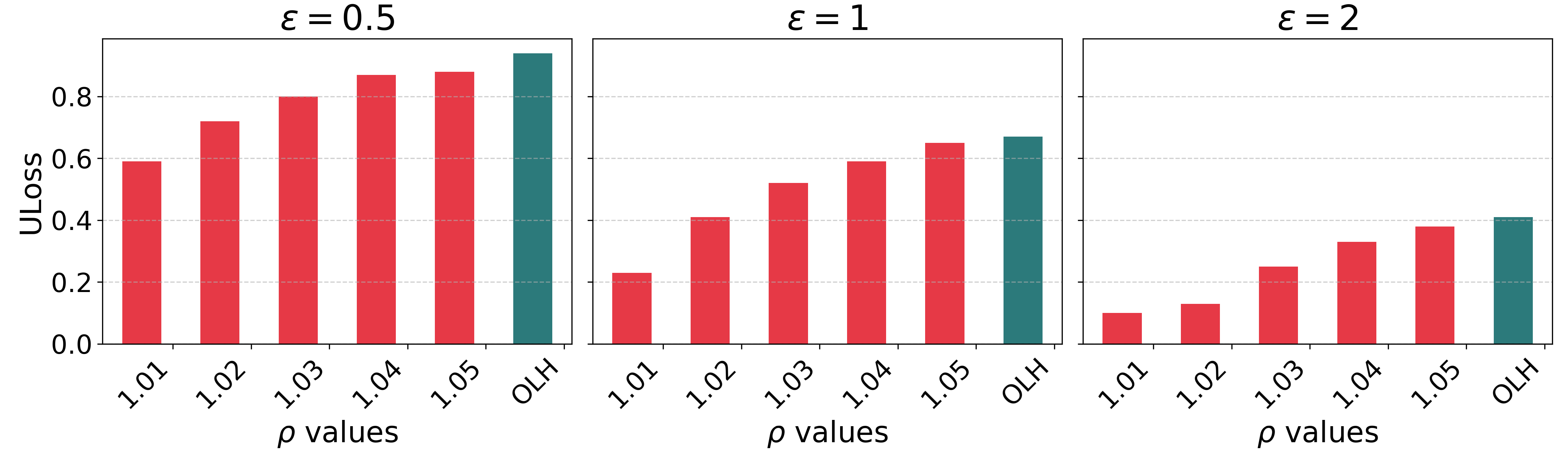}
    \includegraphics[width=.85\textwidth]{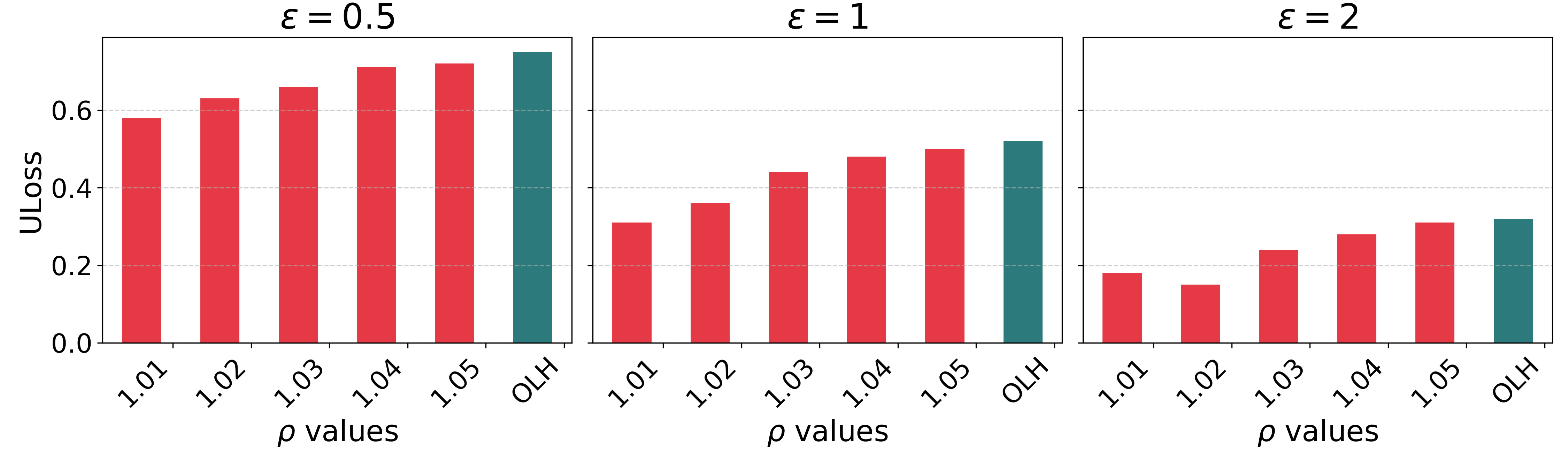}
    \vspace{-10pt}
    \caption{MGA results of the proposed F-OLH protocol using different Gaussian datasets. First row is with $\sigma$ = 1, second row is with $\sigma$ = 5, third row is with $\sigma$ = 10.}
\end{figure}

\section{Selection of the $\rho$ Threshold}

In the paper, we established that selecting $\rho$ close to 1 improves fairness and resilience to both BIA and MGA attacks. However, it also has the adverse effect of increasing execution time. In this section, we provide guidance regarding how to select $\rho$. In particular, when selecting the value of $\rho$, we recommend performing the following analysis, which can be conducted in simulation using $\mathcal{D}$, $\varepsilon$, and $\mathcal{H}$ prior to actual data collection.

For user $u$ with true value $v_u$ and hash function $H_u$, recall that $P_u$ denotes the preimage set. We define:
\begin{equation}
    P_{max} = \max_{u \in \mathcal{U}} |P_u| \qquad P_{min} = \min_{u \in \mathcal{U}} |P_u| \qquad P_{avg} = \frac{1}{|\mathcal{U}|} \sum\limits_{u \in \mathcal{U}} |P_u|
\end{equation}
In Table \ref{tab:rho-selection}, we analyze how different choices of $\rho$ affect $P_{min}$, $P_{max}$, and $P_{avg}$ in F-OLH. We compare these values with theoretical bounds of $P_{min} = 1$, $P_{max} = |\mathcal{D}|$, and $P_{avg} = \frac{|\mathcal{D}|}{g}$, as well as those achieved by OLH. In general, it is desirable for OLH and F-OLH to have $P_{avg}$ similar to the theoretical $P_{avg}$, since it implies higher uniformity of the hash function. On the other hand, it is desirable to have $P_{min}$ and $P_{max}$ that are distant from the theoretical $P_{min}$ and $P_{max}$, since the theoretical $P_{min}$ and $P_{max}$ represent edge cases (maximum non-uniformity).

It can be observed from Table \ref{tab:rho-selection} that as we decrease $\rho$ in F-OLH, the values of $P_{min}$, $P_{max}$, and $P_{avg}$ become more desirable. As we increase $\rho$, the values of $P_{min}$, $P_{max}$, and $P_{avg}$ approach OLH, which is undesirable. It should be noted that even when $\rho$ = 1.01, $P_{avg}$ of F-OLH is different from the theoretical $P_{avg}$, but nevertheless, it is substantially lower than other $\rho$ and OLH. 

\begin{table}[!t]
    \centering
    \renewcommand{\arraystretch}{1.2}
    \caption{$P_{avg}$, $P_{min}$, and $P_{max}$ on BMS-POS and Kosarak datasets, $\varepsilon$ = 2.}
\label{tab:rho-selection}
    \begin{minipage}{0.45\linewidth}
        \centering
        \begin{tabular}{|c|c|c|c|}
            \hline
            & $P_{avg}$ & ~$P_{min}$~ & ~$P_{max}$~ \\
            \hline
            ~Theoretical~ & 12.50 & 1 & 100 \\ \hline
            $\rho = 1.01$ & ~12.86~ & 7 & 19 \\ \hline
            $\rho = 1.02$ & 13.10 & 5 & 22 \\ \hline
            $\rho = 1.03$ & 13.26 & 4 & 25 \\ \hline
            $\rho = 1.04$ & 13.34 & 3 & 26 \\ \hline
            $\rho = 1.05$ & 13.35 & 2 & 27 \\ \hline
            OLH & 13.37 & 2 & 31 \\
            \hline
        \end{tabular}
    \end{minipage}
    \hspace{10pt}
    \begin{minipage}{0.45\linewidth}
        \centering
        \begin{tabular}{|c|c|c|c|}
            \hline
            & $P_{avg}$ & ~$P_{min}$~ & ~$P_{max}$~ \\
            \hline
            ~Theoretical~ & ~16.00~ & 1 & 128 \\ \hline
            $\rho = 1.01$ & 16.30 & 9 & 24 \\ \hline
            $\rho = 1.02$ & 16.58 & 6 & 26 \\ \hline
            $\rho = 1.03$ & 16.64 & 5 & 31 \\ \hline
            $\rho = 1.04$ & 16.71 & 4 & 33 \\ \hline
            $\rho = 1.05$ & 16.74 & 3 & 35 \\ \hline
            OLH & 16.74 & 3 & 36 \\
            \hline
        \end{tabular}
    \end{minipage}
\end{table}

In general, the analysis in Table \ref{tab:rho-selection} can be used in $\rho$ selection as follows. First, one should select $\rho$ such that $P_{avg}$ of F-OLH is not too different from the theoretical $P_{avg}$. Second, considering that $P_{min}$ and $P_{max}$ represent the unfair edge cases, the two should be as similar to one another as possible, e.g., the difference between them should be low. Using $\mathcal{H}$ and $\mathcal{D}$, the data collector can simulate $P_{min}$, $P_{max}$, and $P_{avg}$ under varying $\rho$. Similar to Table \ref{tab:rho-selection}, these values can be compared with their theoretical values and the outputs of OLH. The $\rho$ value that yields sufficient divergence from OLH and similarity to the theoretical $P_{avg}$ can be selected as the $\rho$ to be used. It should be noted that as $\rho$ gets smaller (i.e., stricter), it will be more difficult to find a hash function that satisfies this $\rho$, and the execution time will increase.

\end{document}